\documentclass [a4paper]{PoS}
\newcommand{\agev}    {\mbox{$A$~GeV}}               

\newcommand{\rb}[1]   {\mbox{\textrm{\scriptsize #1}}}
\newcommand{\rbt}[1]  {\mbox{\textrm{\tiny #1}}}
\newcommand{\sqrts}   {\ensuremath{\sqrt{s_{_{\rbt{NN}}}}}}

\newcommand{\lam}     {\ensuremath{\Lambda}}
\newcommand{\lab}     {\ensuremath{\bar{\Lambda}}}  
\newcommand{\xim}     {\ensuremath{\Xi^{-}}}
\newcommand{\xizero}  {\ensuremath{\Xi^{0}}}
\newcommand{\xip}     {\ensuremath{\bar{\Xi}^{+}}}

\newcommand{\nwound}  {\ensuremath{\langle N_{\rb{w}} \rangle}}

\newcommand{\dnetbar} {\ensuremath{\textrm{d}N_{(\rbt{B}-\bar{\rbt{B}})}/\textrm{d}y}}

\title{Recent results on (anti)nucleus and (anti)hyperon production 
       in nucleus-nucleus collisions at CERN SPS energies}

\ShortTitle{(Anti)nucleus and (anti)hyperon production in A+A at CERN SPS}

\author{\speaker{Georgy Melkumov} for the NA49 Collaboration \\

        Joint Institute for Nuclear Research, Dubna\\
        E-mail: \email{Georgui.Melkoumov@cern.ch}}

\author{The NA49 Collaboration: \\
C.~Alt$^{9}$, T.~Anticic$^{23}$, B.~Baatar$^{8}$,D.~Barna$^{4}$,
J.~Bartke$^{6}$, L.~Betev$^{10}$, H.~Bia{\l}\-kowska$^{20}$,
C.~Blume$^{9}$,  B.~Boimska$^{20}$, M.~Botje$^{1}$
% ...
}

\author{ \\
$^{1}$NIKHEF, Amsterdam, Netherlands. \\
% ...
$^{4}$KFKI Research Institute for Particle and Nuclear Physics, Budapest, Hungary.\\
% ...
$^{6}$Henryk Niewodniczanski Institute of Nuclear Physics, Polish Academy of Sciences, Cracow, Poland.\\
% ...
$^{8}$Joint Institute for Nuclear Research, Dubna, Russia.\\
$^{9}$Fachbereich Physik der Universit\"{a}t, Frankfurt, Germany.\\
$^{10}$CERN, Geneva, Switzerland.\\
% ...
$^{20}$Institute for Nuclear Studies, Warsaw, Poland.\\
% ...
$^{23}$Rudjer Boskovic Institute, Zagreb, Croatia.\\
% ...

}

\abstract{

The NA49 experiment has collected comprehensive data on particle production in 
nucleus-nucleus collisions over the whole SPS beam energies range, the critical 
energy domain where the expected phase transition to a deconfined phase is expected 
to occur.
The latest results from Pb+Pb collisions between 20$A$ GeV and 158$A$ GeV on baryon 
stopping and light nuclei production as well as those for strange hyperons are 
presented.  
  
The measured data on $p$, $\bar{p}$, $\Lambda$, $\bar{\Lambda}$, $\Xi^-$ and 
$\bar{\Xi}^+$ production were used to evaluate the rapidity distributions of net-baryons 
at SPS energies and to compare with the results from the AGS and the RHIC for central 
Pb+Pb (Au+Au) collisions.

The dependence of the yield ratios and the inverse slope parameter of the $m_t$
spectra on the collision energy and centrality, and the mass number of the
produced nuclei $^3He$, $t$, $d$ and $\bar{d}$ are discussed within coalescence 
and statistical approaches.
Analysis of the total multiplicity exhibits remarkable agreement between the
measured yield for $^3He$ and those predicted by the statistical hadronization
model.
    
   In addition, new results on $\Lambda$ and $\bar{\Lambda}$ as well as $\Xi^-$
production in minimum bias Pb+Pb reactions at 40$A$ GeV and 158$A$ GeV and central 
C+C, Si+Si and Pb+Pb collisions are presented.
  The system size dependence of the yields of these hyperons was analysed to 
determine the evolution of strangeness enhancement relative to elementary p+p
collisions.
 }

\FullConference{Critical Point and Onset of Deconfinement
          4th International Workshop\\
		 July 9-13 2007\\
		 GSI Darmstadt,Germany}

\begin{document}

\section{Introduction}

The NA49 experiment studied high energy and density matter produced in 
nucleus-nucleus collisions at the CERN SPS to search for evidence of 
quark-gluon deconfinement in the early stage of the reactions predicted 
by lattice QCD. 
In the first measurements NA49 demonstrated that in central Pb+Pb collisions 
top SPS energy the initial energy density well exceeds the critical 
value of about 1 GeV/$fm^3$. The results of the energy scan measurements 
between 20$A$ GeV and 158$A$ GeV beam energies suggest that deconfinement indeed 
starts in the lower SPS energy range \cite{gazd2004,zak2006}. Among a primary 
features, the SPS experiments found that the relativistic heavy 
ion reactions produce an explosively expanding fireball with strong transverse 
and longitudinal flow. 
Another characteristics is that the ratios of yields of produced particles are 
consistent with statistical equilibration.

In this report a new results related to the formation of hot and dense nuclear 
matter in $A-A$ collisions and its dynamics - energy dependence of baryon 
stopping and light nucleus production, and possible evidence of deconfinement 
- size dependence of strangeness enhancement will be presented.

\section{Experiment}

The experiment was carried out with the NA49 large acceptance hadron detector 
\cite{afan1999} employing a system of time projection chambers (TPCs) for efficient 
tracking in the forward hemisphere of the reactions, precise momentum reconstruction 
in the magnetic field and particle identification using the energy loss $dE/dx$ in 
the TPC gas.
Two time of flight (TOF) walls of 900 scintillator pixels each situated symmetrically 
behined TPCs augment particle identification near midrapidity.
A zero degree calorimeter ZDC is used for centrality selection of events in nucleus
-nucleus collisions at triggering as well as in the off-line analysis.  

Raw charged particle yields were obtained by unfolding the distributions
of $dE/dx$ and from $TOF$ measurements in small bins of momentum $p$ and transverse
momentum $p_t$ \cite{afan2002,ppbarPRC2006}.
Strange particles ($K^0_S$, $\Lambda$, $\Xi$, $\Omega$) are detected via decay topology 
and invariant mass measurement \cite{anti2004,alt2005}. 
A various type of corrections have been applied to the data, such as corrections for 
geometrical acceptance, reconstruction efficiency, in-flight particle decay and 
feeddown from weak decays.

\begin{figure}[htb]
\begin{center}
\begin{minipage}[b]{35mm}
\begin{center}
\includegraphics[height=45mm]{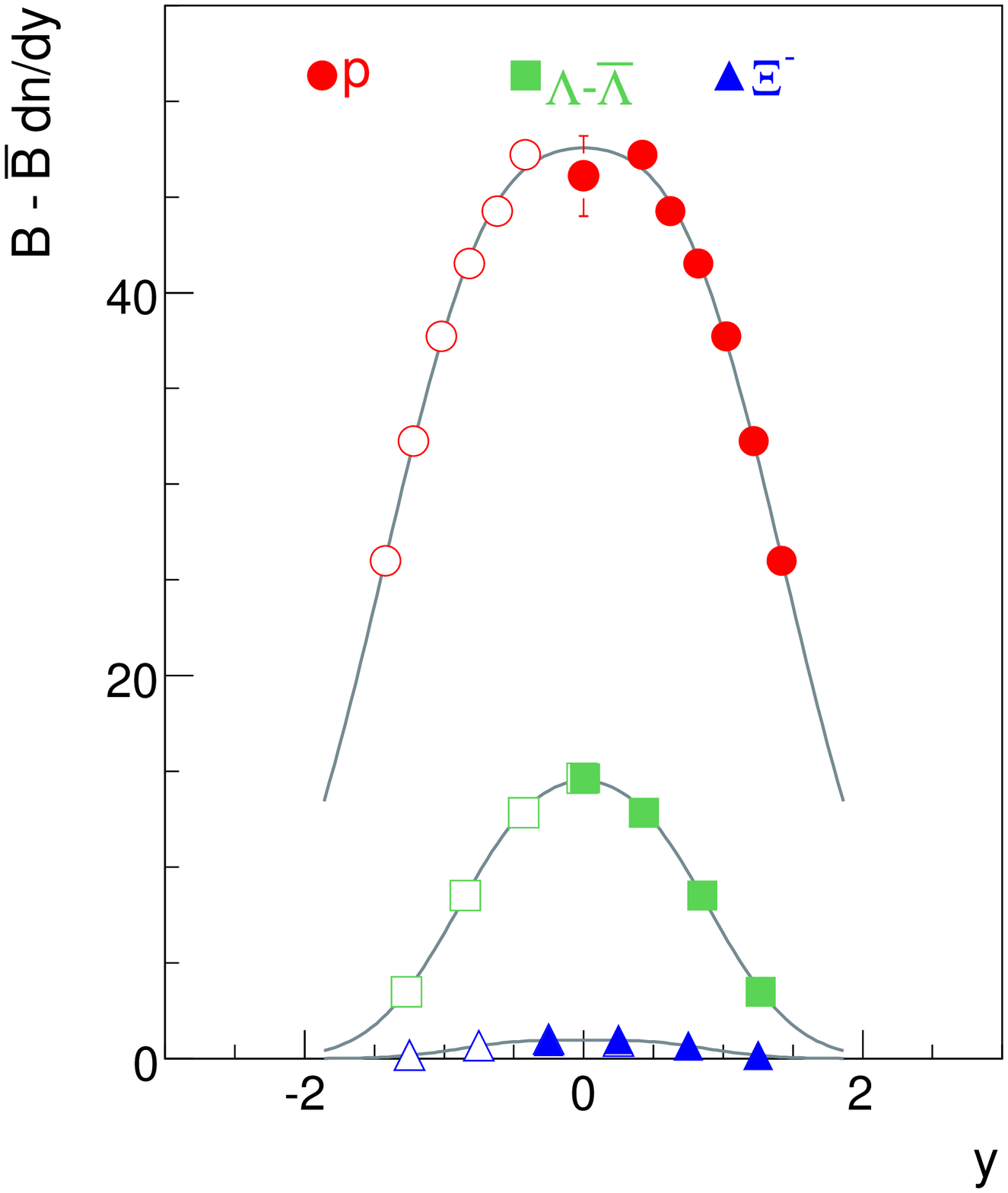}
\end{center}
\end{minipage}
\begin{minipage}[b]{35mm}
\begin{center}
\includegraphics[height=45mm]{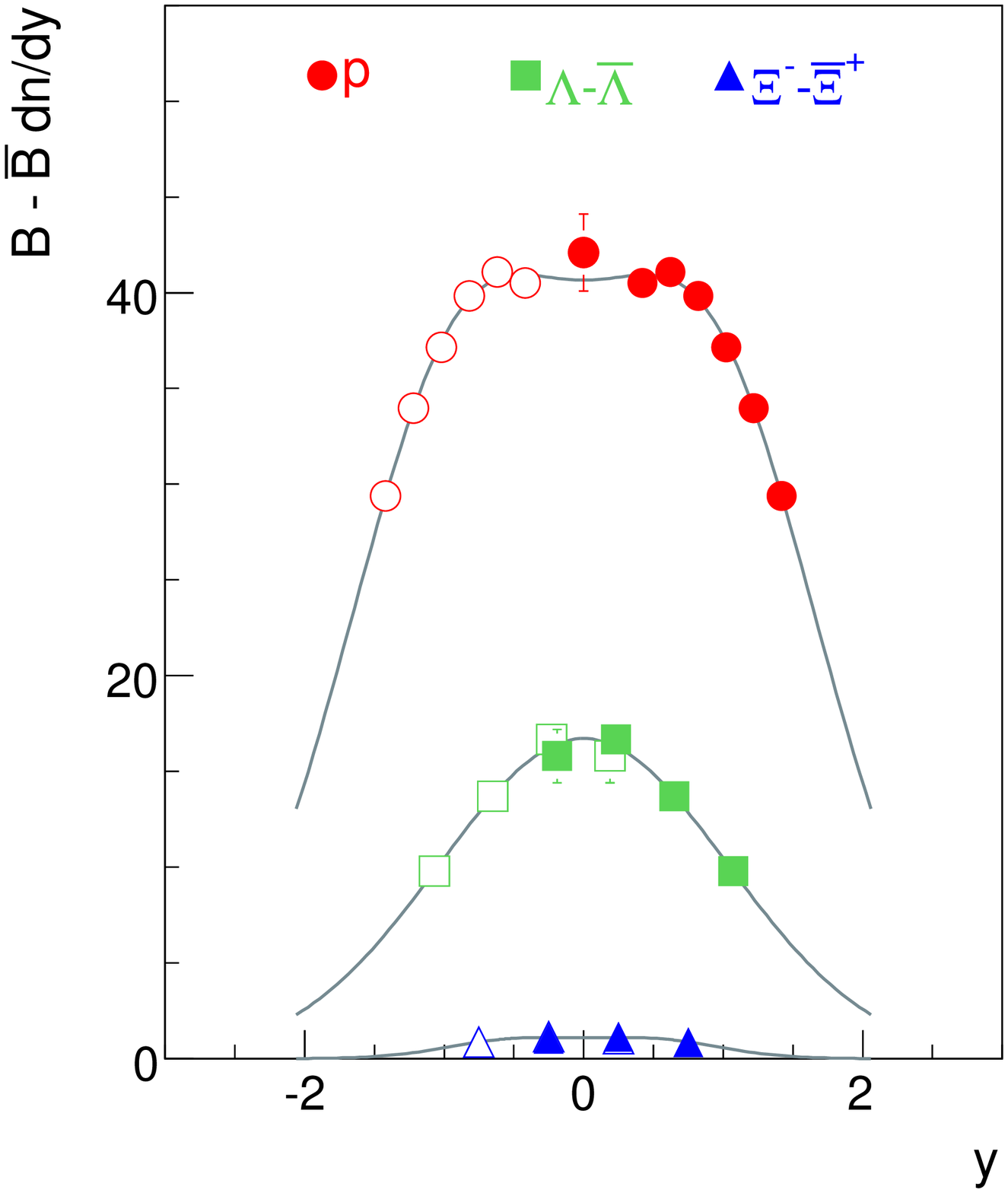}
\end{center}
\end{minipage}
\begin{minipage}[b]{35mm}
\begin{center}
\includegraphics[height=45mm]{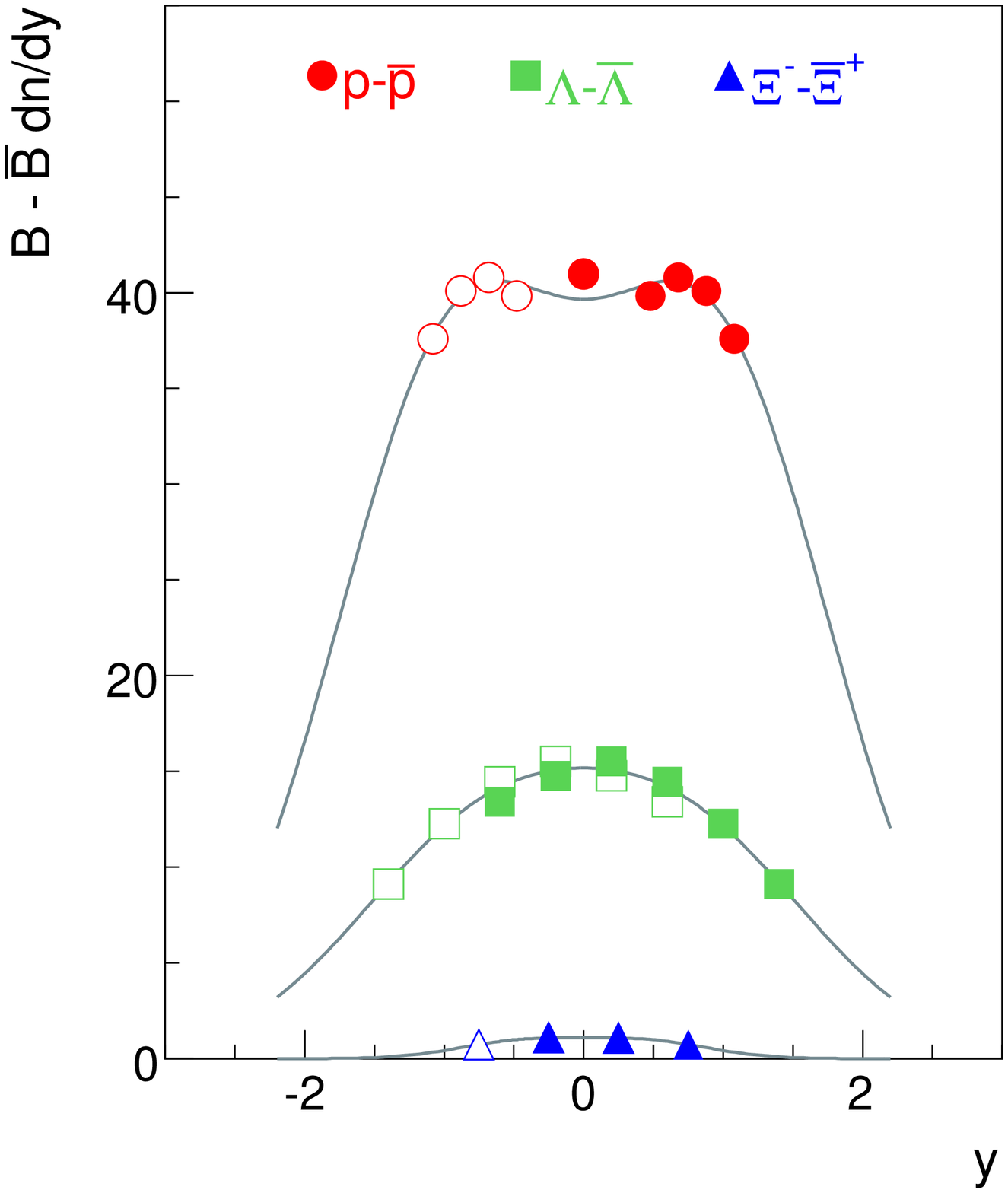}
\end{center}
\end{minipage}
\begin{minipage}[b]{35mm}
\begin{center}
\includegraphics[height=45mm]{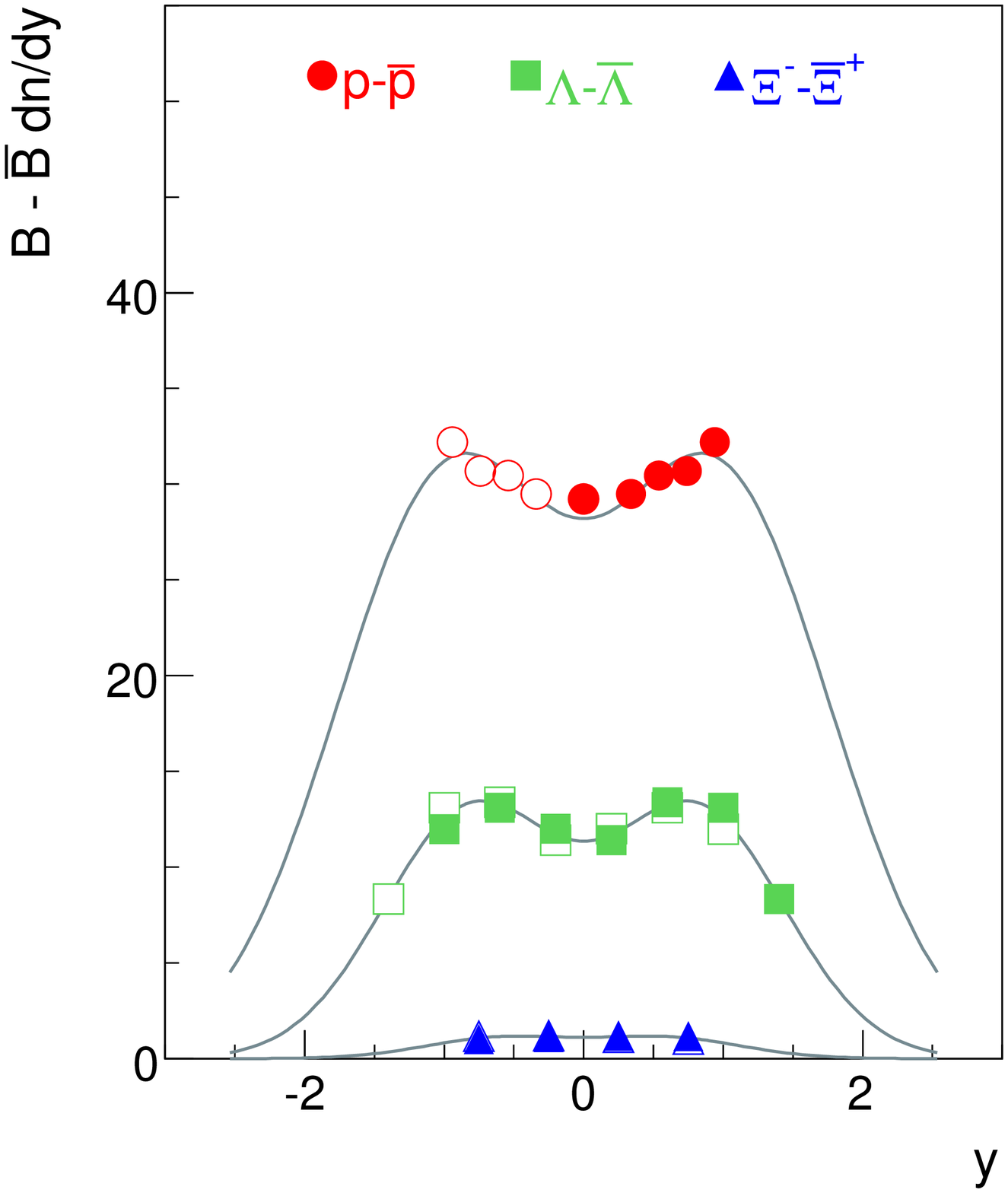}
\end{center}
\end{minipage}
\end{center}
\caption{Rapidity distributions of net baryons $p$, $\Lambda$ and $\Xi$ 
measured in the most 7\% cemtral Pb+Pb collisions at CERN SPS energies 
(left to right) 20$A$, 30$A$, 40$A$, 80$A$~GeV. Open data points were 
obtained by reflection at midrapidity.}
\label{fig1}
\end{figure}

\section{Baryon stopping}

It is important in relativistic heavy ion experiments to understand nuclear 
stopping in details since it is a requisite for the formation of hot and dense 
nuclear matter. 
Stopping is a necessary ingradient in our overall understanding of the reaction 
and the expectations to form and study the properties of QGP.

\begin{figure}[htb]
\begin{center}
\begin{minipage}[b]{50mm}
\begin{center}
\includegraphics[height=47mm]{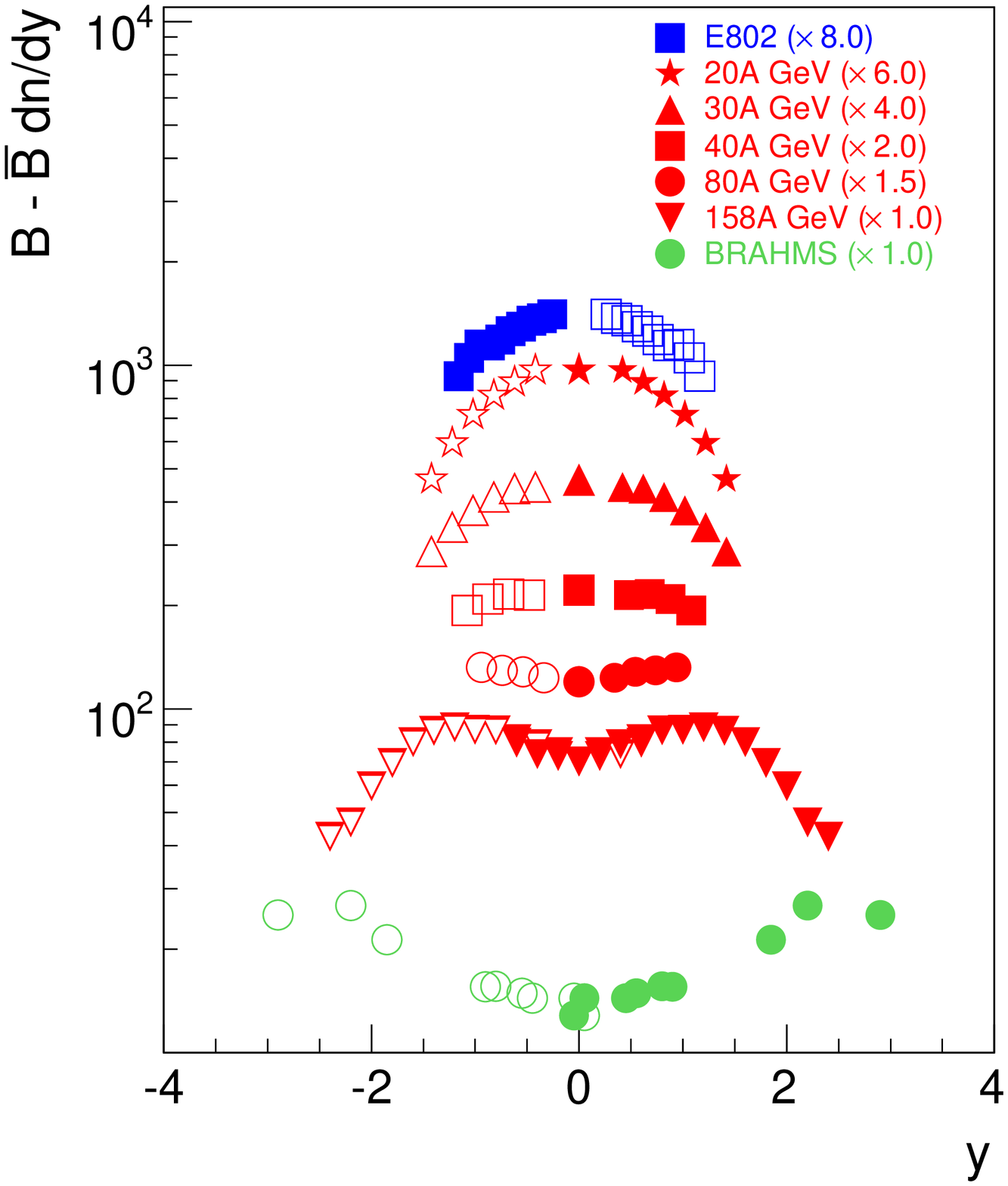}
\end{center}
\end{minipage}
\begin{minipage}[b]{50mm}
\begin{center}
\includegraphics[height=47mm]{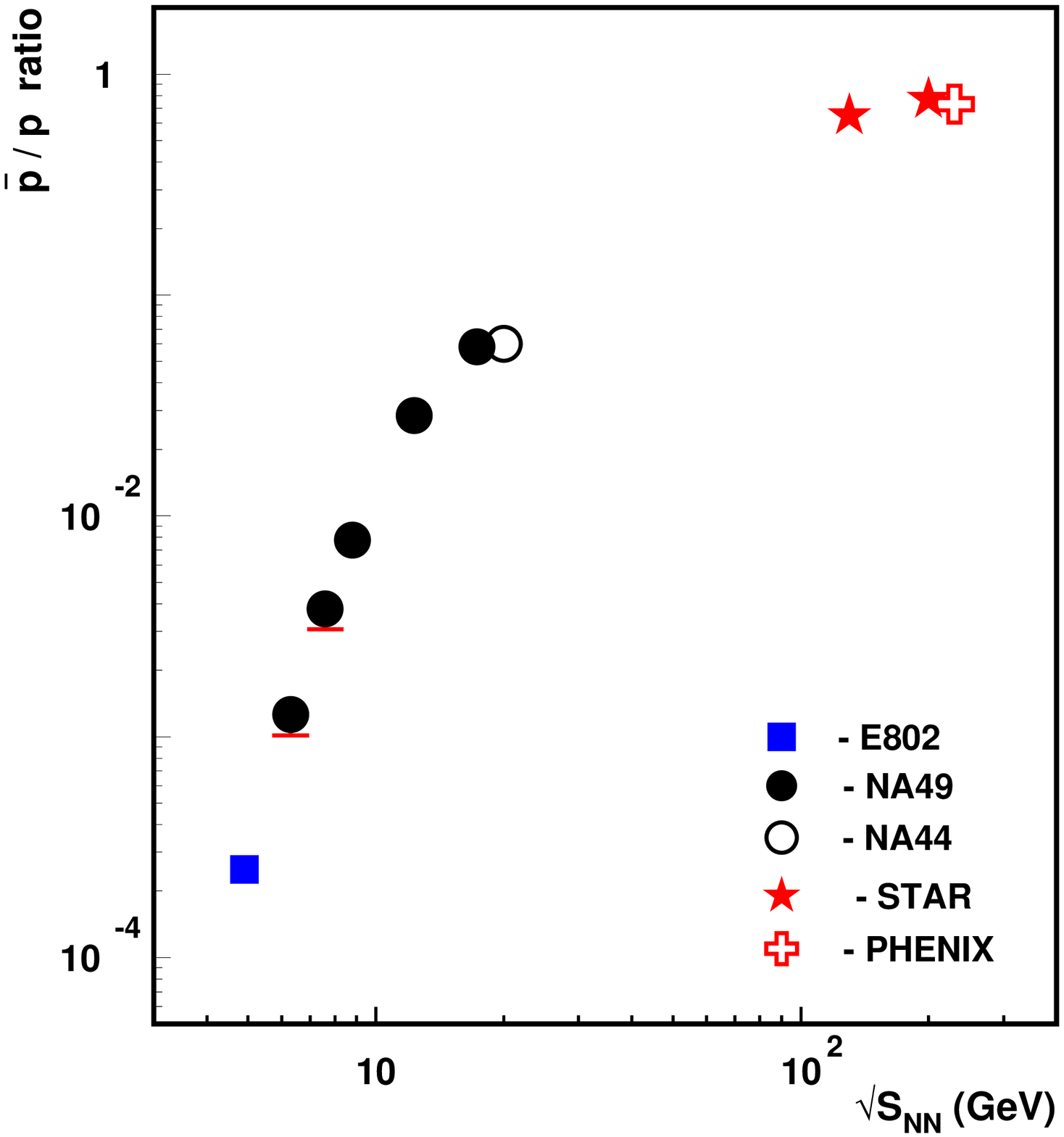}
\end{center}
\end{minipage}
\begin{minipage}[b]{45mm}
\begin{center}
\includegraphics[height=45mm]{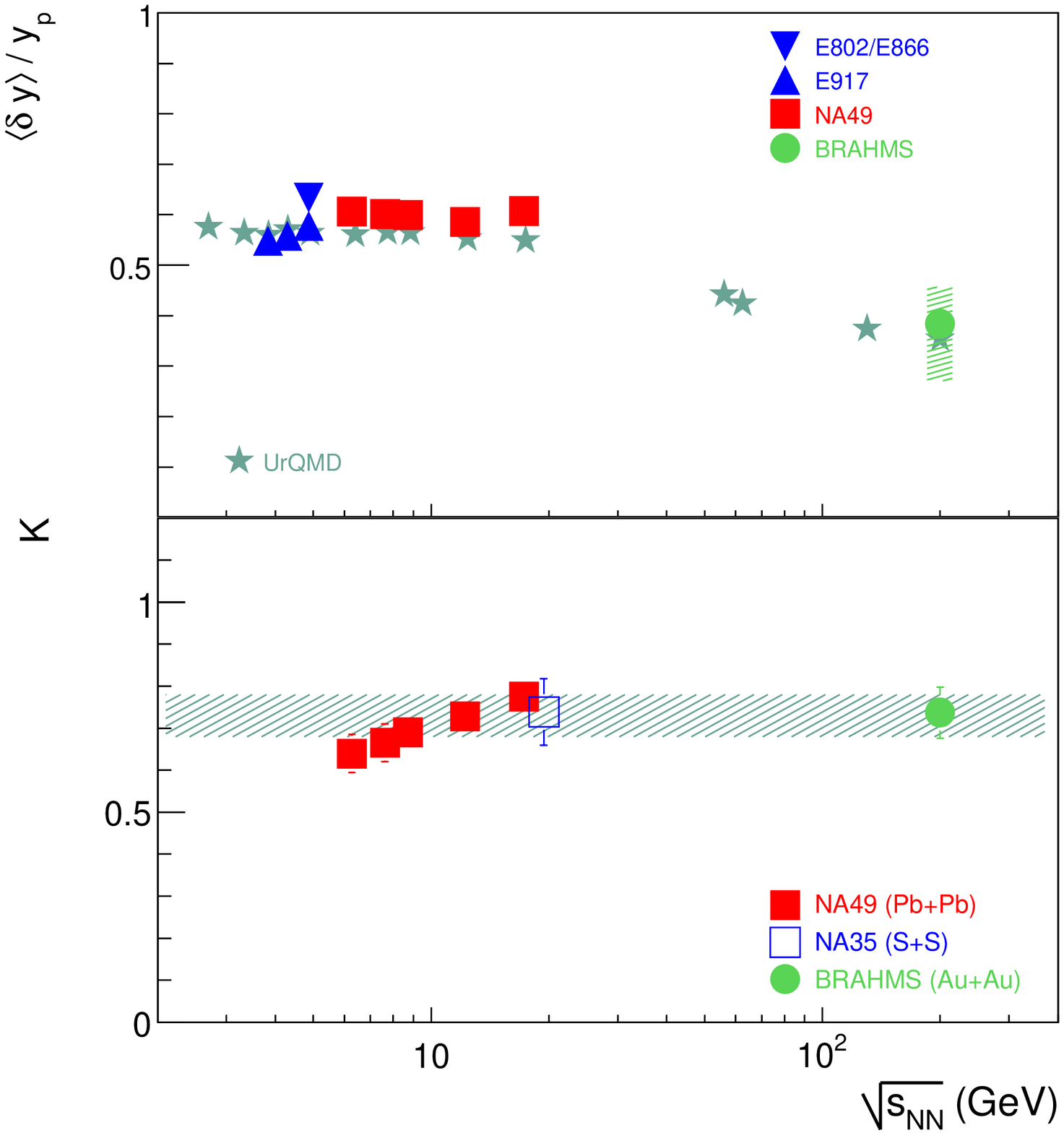}
\end{center}
\end{minipage}
\end{center}
\caption{
(left) Net baryon rapidity distributions at AGS\cite{ahle1999}, 
SPS\cite{appel1999} and RHIC\cite{beard2004} energies in central Pb+Pb 
or Au+Au collisions. Open data points were obtained by reflection. 
(center) $\bar{p}/p$ ratio at midrapidity as a function of $\sqrt{s_{NN}}$ 
in central Pb+Pb (Au+Au) collisions reviwed in \cite{ppbarPRC2006}.
(right top) Relative rapidity shifts of projectile nucleons extracted from net
baryon distributions compared to results (stars) from the UrQMD model 
calculations \cite{urqmd}.
(right bottom) Dependence of the inelasticity factor $K$ on $\sqrt{s_{NN}}$. 
}
\label{fig2}
\end{figure}

New results on rapidity spectra of protons and antiprotons in central Pb+Pb
reactions at 20$A$ - 80$A$ GeV shown in Fig.~\ref{fig1} in combination with 
previously published results \cite{appel1999,ppbarPRC2006} allow to study the 
energy evolution of stopping.   
Based on the measured rapidity spectra for $p$, $\bar{p}$ 
\lam, \lab, \xim and \xip (Fig.~\ref{fig1}), all corrected for feed down
from weak decays, the net-baryon distributions \dnetbar\ are constructed:  
\begin{equation}
 B - \bar{B} = S_n \cdot ( p - \bar{p} ) 
       + S_{\Sigma^{\pm}} \cdot (\Lambda - \bar{\Lambda})
       + S_{\Xi^0} \cdot (\Xi^- - \bar{\Xi}^{+} ) . 
\label{eq1}
\end{equation}
The contribution of unmeasured baryons (n, $\Sigma^{\pm}$, \xizero) was 
estimated using a scaling factors $S_x$ obtained from statistical hadron 
gas model fits \cite{becc2007}.
The results are plotted in Fig.~\ref{fig2}~(left). A clear evolution is seen
from a peak to a dip structure in the SPS energy range. 

The averaged rapidity shift of projectile nucleons $\langle \delta y \rangle$, 
which is commonly used to quantify stopping in $A-A$ collisins, can be derived 
from:
\begin{equation}
\langle \delta y \rangle = y_{proj} - \frac{2}{N_{part}}
 \int_0^{y_{proj}} y \frac{dN_{B-\bar{B}}}{dy} dy ,
\label{eq2}
\end{equation}
where $y_{proj}$ is the projectile rapidity and $N_{\rb{part}}$ is the number
of participating nucleons.  
As seen from Fig.~\ref{fig2}~(upper right), $\langle \delta y \rangle$ is 
approximately 0.6 at AGS and SPS and then slowly decreases in the RHIC energy 
range. This behaviour is well reproduced by the UrQMD model 
calculations \cite{urqmd}.  

Another measure promptly related to the stopping power is the $\bar{p}$/$p$
ratio and its dependence on the energy and centrality of the interactions.   
Fig.~\ref{fig2}~(center) displays the midrapidity $\bar{p}$/$p$ ratio for 
central collisions as a function of energy at the AGS, SPS and RHIC early 
reviewed in \cite{ppbarPRC2006}.
The ratio rises steeply within the SPS energy range by nearly two orders
of magnitude. The figure illustrates how the collisions evolve from producing a
net baryon-rich system at the AGS through the SPS energy range to an almost net 
baryon-free midrapidity region at the RHIC. 

The partial stopping of incident nucleons observed in $A$-$A$ collisions 
provides the energy for particle production, which can be estimated in terms 
of the inelasticity. 
Using \dnetbar\ and the measured $\langle m_{\rb{t}} \rangle$ the inelastic energy 
per net-baryon:
\begin{equation}
E_{\rb{inel}} = \frac{\sqrt{s_{\rbt{NN}}}}{2} - 
                \frac{1}{N_{(\rbt{B}-\bar{\rbt{B}})}}
                \int_{-y_{proj}}^{y_{proj}} \langle m_{\rb{t}} \rangle \:
                \frac{\textrm{d}N_{(\rbt{B}-\bar{\rbt{B}})}}{\textrm{d}y} 
                \: \cosh y \: \textrm{d}y
\end{equation}
and the inelasticity factor $K = 2 \:E_{\rb{inel}} / (\sqrt{s_{\rbt{NN}}} - 
2 m_{\rb{p}})$ can be calculated.  
Fig.~\ref{fig2}~(lower right) indicates that $K$ is approximatly energy independent 
and the average energy loss of projectiles amounts to about 70 - 80 \% at all energies.

\section{Production of light nuclei}\label{lnuclei}

There is a notion that abandances of light nuclei probe the latest stage 
of the evolution of a systems created in relativistic heavy ion collision. 
After the produced system has cooled and expanded the nucleons in close proximity and 
moving with small relative momenta coalesce to form nuclei. Thus light nuclei
production enables the study of several topics, including the mechanism of 
composite particle production, freeze-out temperature, size of the interaction
region and enrtopy of the system. 
Production of light nuclei has traditionally been interpreted in terms of 
coalescence models \cite{poll1998,heinz1999}. 

Recently NA49 has measured the energy dependence of deuteron snd tritons 
production at midrapidity, and $^3He$ nuclei in a wide rapidity range.  
A rare process of the antideuteron production at midrapidity in central Pb+Pb
collisions at 158$A$ GeV beam energy was also studied. It is of significant 
interest because the initial colliding system contains no antibaryons therefore 
their yields and spectra are 
determined solely by the post collision dynamics. 
It is also reasonable in the antideuteron analysis to assume that the ratio of 
$\bar{d}$ to $\bar{p}$ prior to coalescence is close to 1, whereas for deuterons the 
 coalessence probability will depend upon the beam and target nuclei, 
 as well as  other details of the reaction.
Additionally, the produced antideuterons have no contribution from spectator 
fragments.

 \begin{figure}[htb]
\begin{center}
\begin{minipage}[b]{48mm}
\begin{center}
\includegraphics[height=45mm]{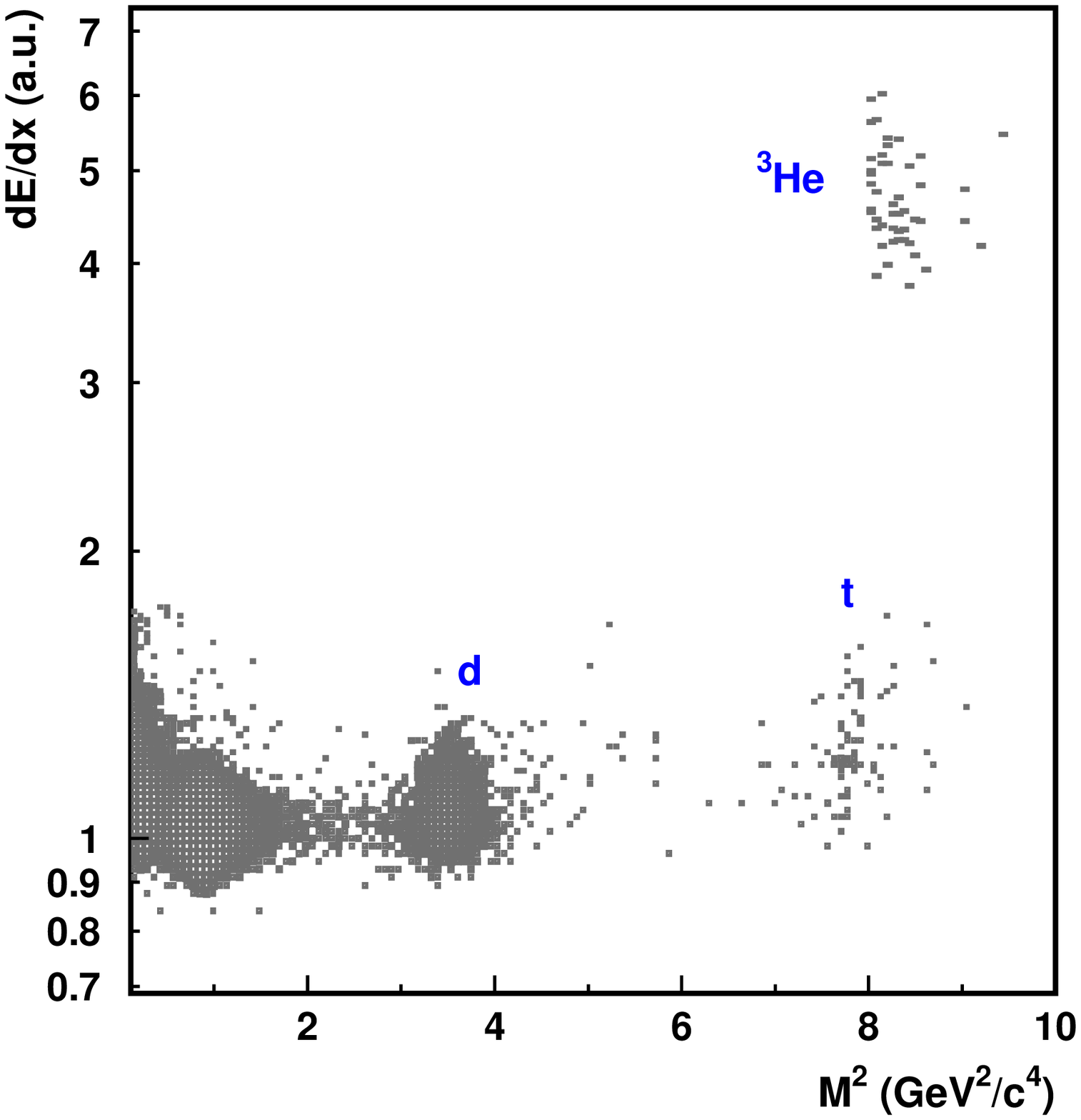}
\end{center}
\end{minipage}
\begin{minipage}[b]{48mm}
\begin{center}
\includegraphics[height=42mm]{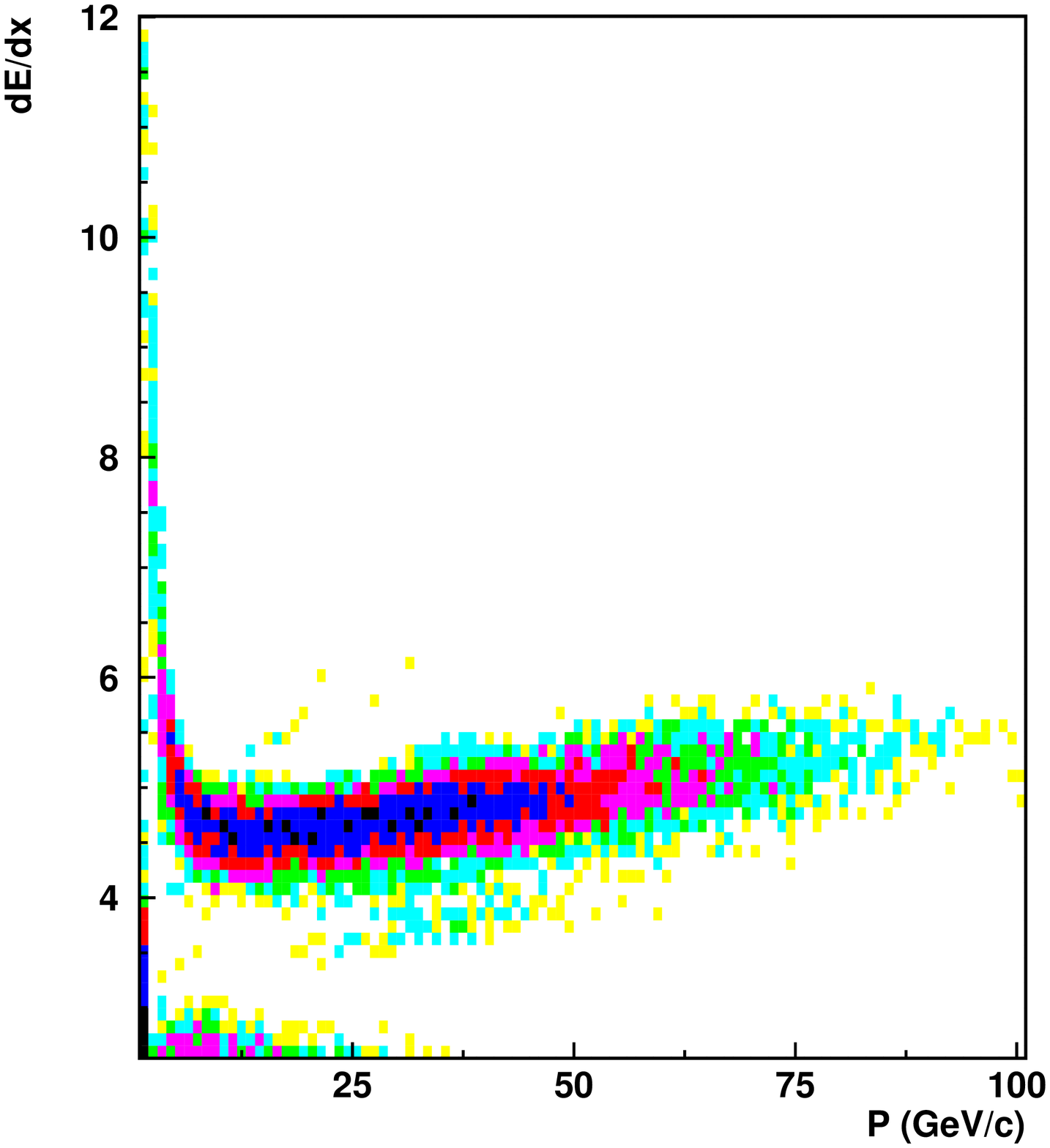}
\end{center}
\end{minipage}
\begin{minipage}[b]{48mm}
\begin{center}
\includegraphics[height=42mm]{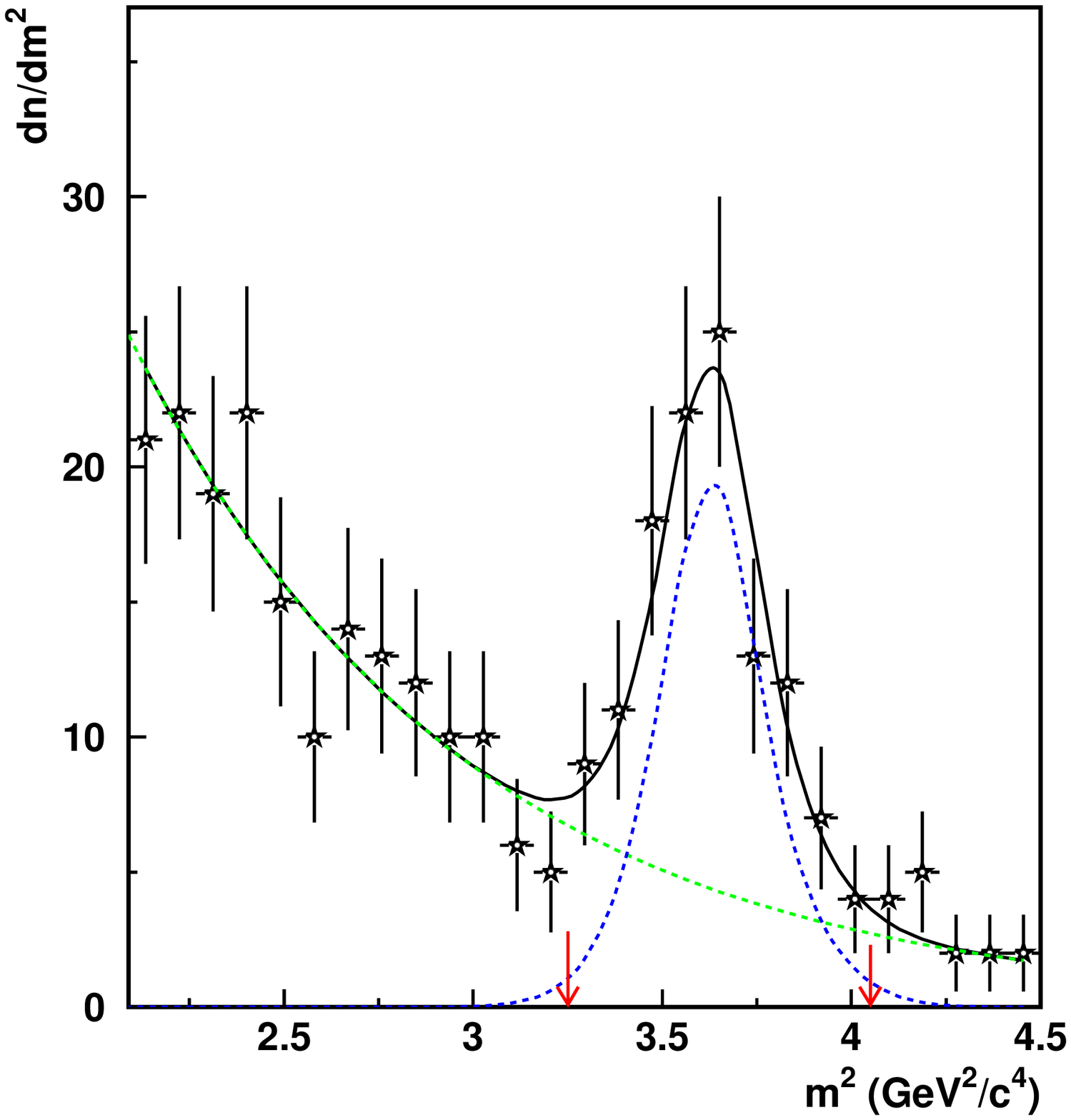}
\end{center}
\end{minipage}
\end{center}
\caption{ 
(left) Energy loss $dE/dx$ in TPC versus mass squared $M^2$ from measurements  
      in TOF for $^3He$, $t$, and $d$ produced in central Pb+Pb collisions. 
(center) Momentum dependence of the $^3He$ energy loss $dE/dx$. 
(right) Mass squared distribution for $\bar{d}$ after $dE/dx$ and $TOF$ 
       identication cuts.
}
\label{fig3}
\end{figure} 
 
 In such investigations the identification of the produced nuclei is of
 primary importance. The Fig.~\ref{fig3}~(left) demonstrates the clean 
 identification of $d$, $t$ and $^3He$ from measurements of the energy loss 
 $dE/dx$ in the TPC gas and mass squared $M^2$ computed from momentum and time 
 of flight measurements in TOF detector (see \cite{dpPRC2004} for more details). 
As shown in Fig.~\ref{fig3}~(center), for doubly charged $^3He$ nuclei the dE/dx 
measerement was sufficient to uniquely and cleanly determine its identity 
thus providing a large coverage in rapidity for measuring $^3He$ in TPCs.
For $\bar{d}$ a strong $dE/dx$ and $TOF$ cuts were required to provide a reliable 
particle identification. 
The  $M^2$ distribution after all identification cuts is shown on 
Fig.~\ref{fig3}~(right).
Antideuterons are clearly visible at about $M^2\approx{3.5}$ GeV$^2/c^4$.

\begin{figure}[htb]
\begin{center}
\begin{minipage}[b]{45mm}
\begin{center}
\includegraphics[height=45mm]{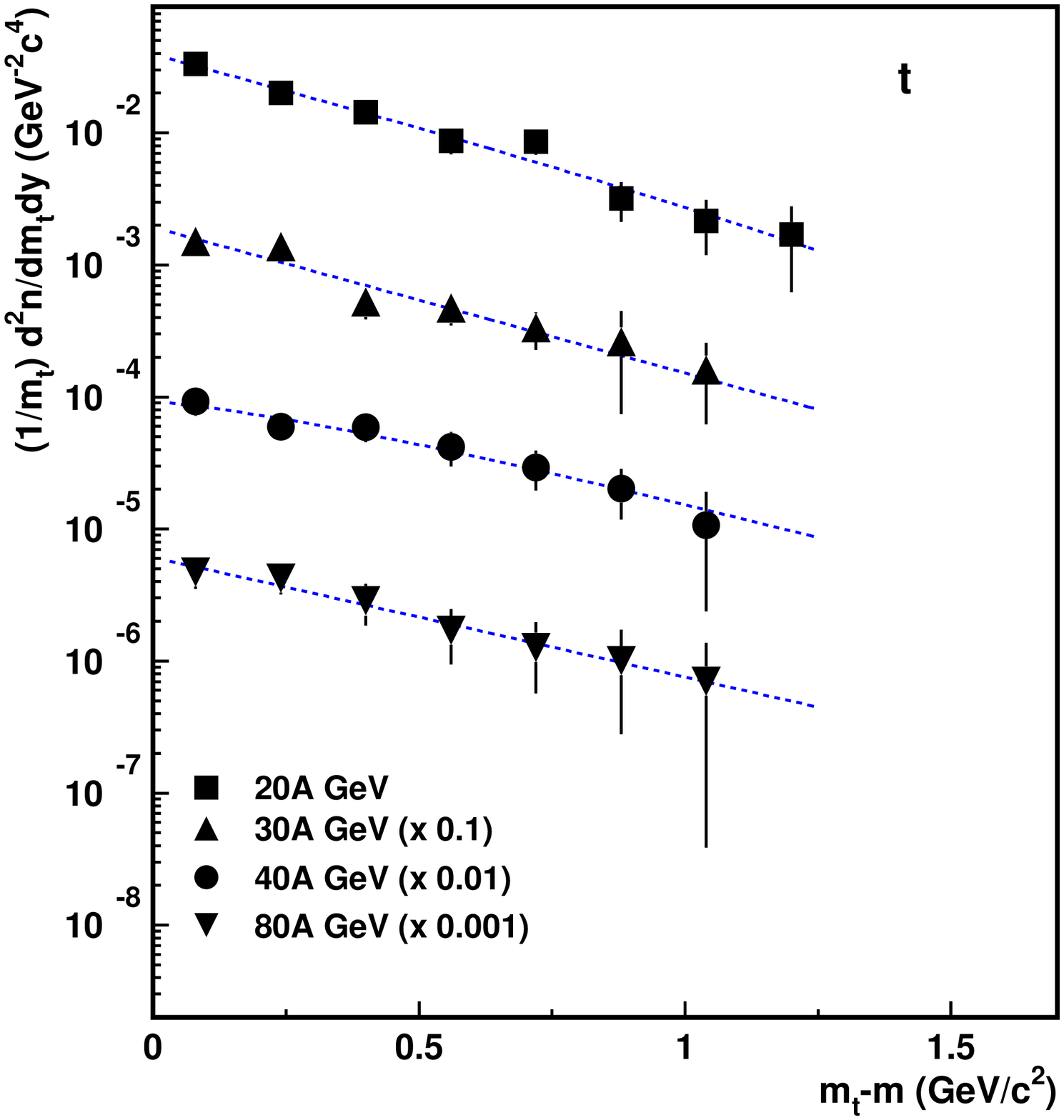}
\end{center}
\end{minipage}
\begin{minipage}[b]{45mm}
\begin{center}
\includegraphics[height=45mm]{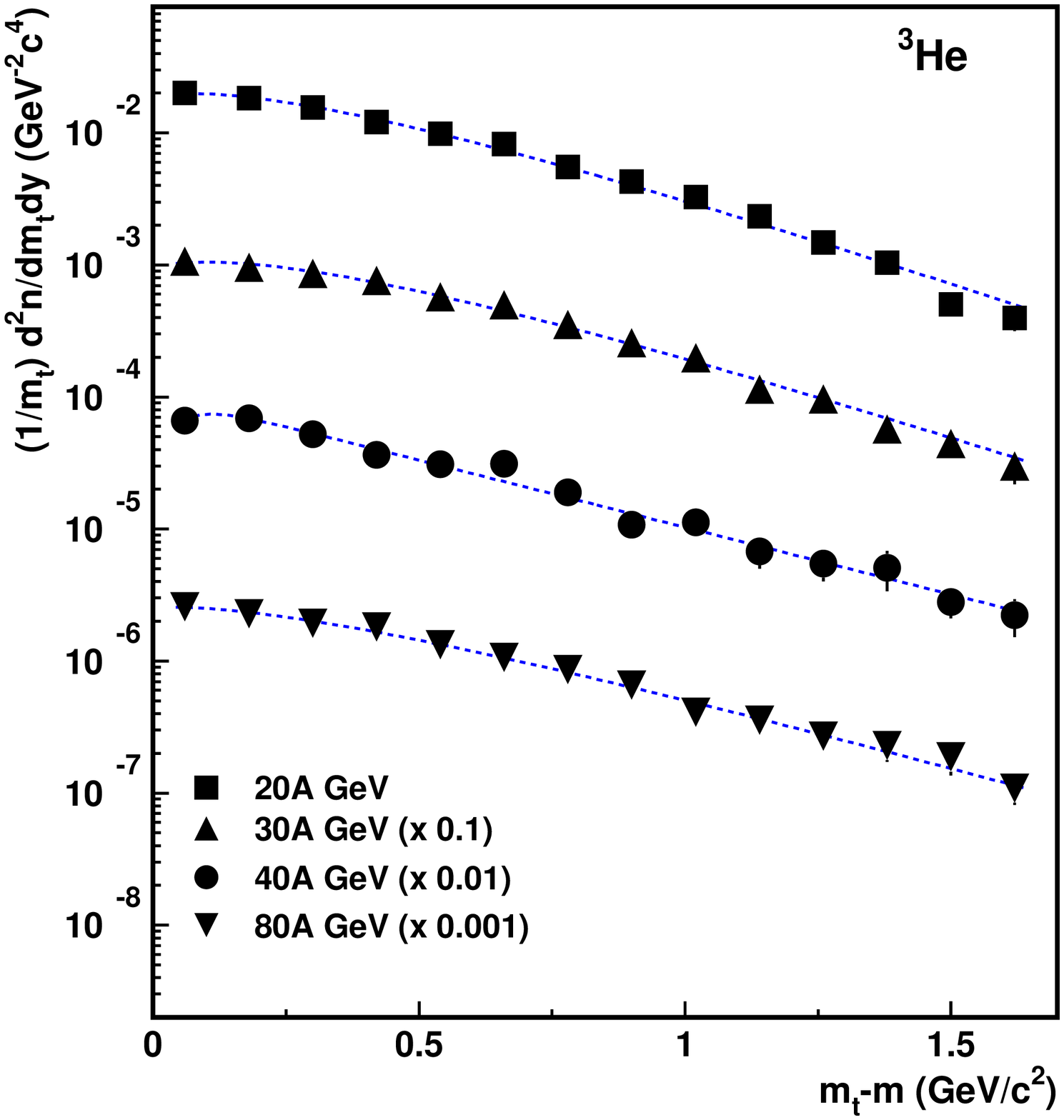}
\end{center}
\end{minipage}
\begin{minipage}[b]{45mm}
\begin{center}
\includegraphics[height=48mm]{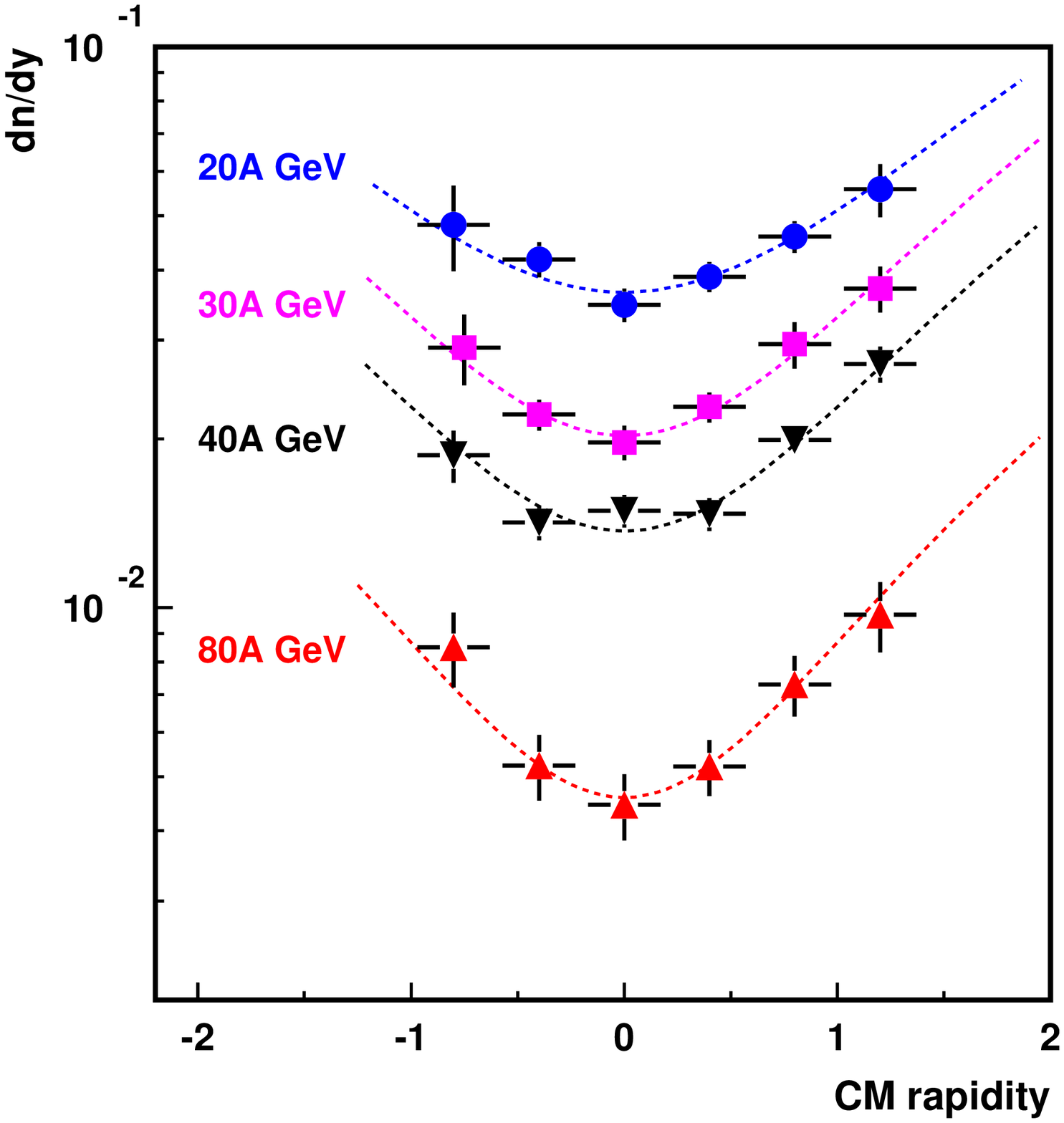}
\end{center}
\end{minipage}
\end{center}
\caption{
Transverse mass $m_t$ distributions of $t$ (left) and $^3He$ (center) 
at midrapidity in the 7\% most central Pb+Pb collisions at 20$A$, 30$A$, 
40$A$ and 80 \agev beam energies. 
Rapidity distributions of $^3He$ (right). 
The dashed lines are fits 
	with a parabola used to extract the total multiplicities.
}
\label{fig4}
\end{figure}

New results on triton and $^3He$ production in central Pb+Pb collisions at 
20$A$ - 80$A$~GeV are shown in Fig.~\ref{fig4}. 
The left and central panels display the invariant yields of $t$ and $^3He$ at 
midrapidity as a function of the trasverse mass $m_t$ for all four beam energies. 
Rapidity distributions $dn/dy$, summarised in the right panel of Fig.~\ref{fig4}, 
were obtained by integration of the $m_t$ distributions in small bins of rapidity.
One sees from the figure that the $dn/dy$ distributions have a concave shape 
at all investigated energies.  
This is in constrast to the case of protons, where the shape of rapidity 
distributions is rather convex and strongly energy dependent (see Fig.~\ref{fig1}).  
The rapidity density distributions for light nuclei were studied in \cite{mattiello1997,
hansen1999,rqmd} using the RQMD model in combination with a coalescence model 
applying an "after burner" program to the final proton and neutron distributions.
Spectators were not included in the calculations. 
It was shown that distributions are far from simple thermal Boltzmann distributions 
and the yield of light nuclei increases as one goes from midrapidity towards spectator 
rapidity regions. This effect was explained due to the strong correlation of nucleons 
in the RQMD source.

\begin{figure}[htb]
\begin{center}
\begin{minipage}[b]{45mm}
\begin{center}
\includegraphics[height=48mm]{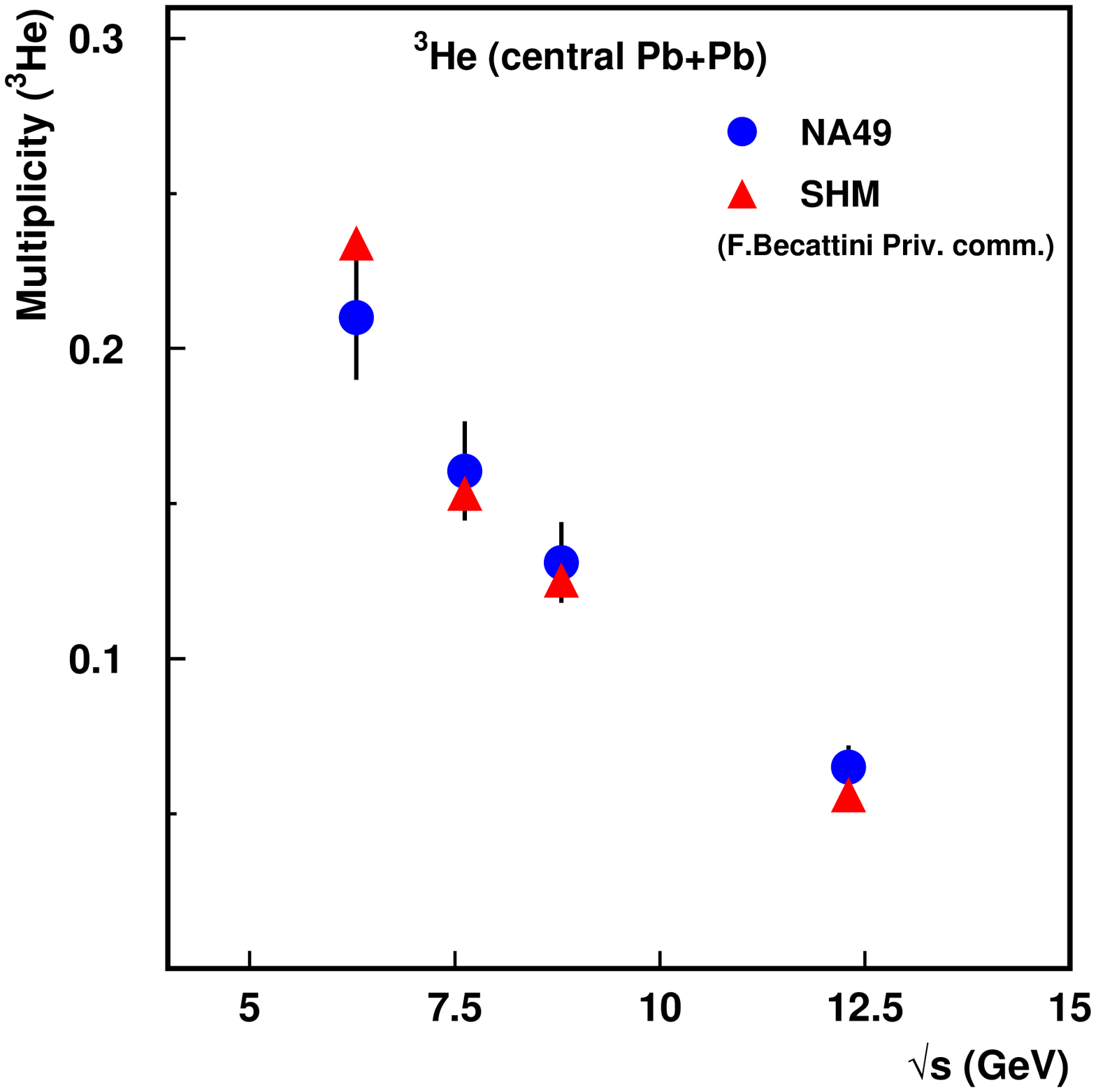}
\end{center}
\end{minipage}
\begin{minipage}[b]{45mm}
\begin{center}
\includegraphics[height=48mm]{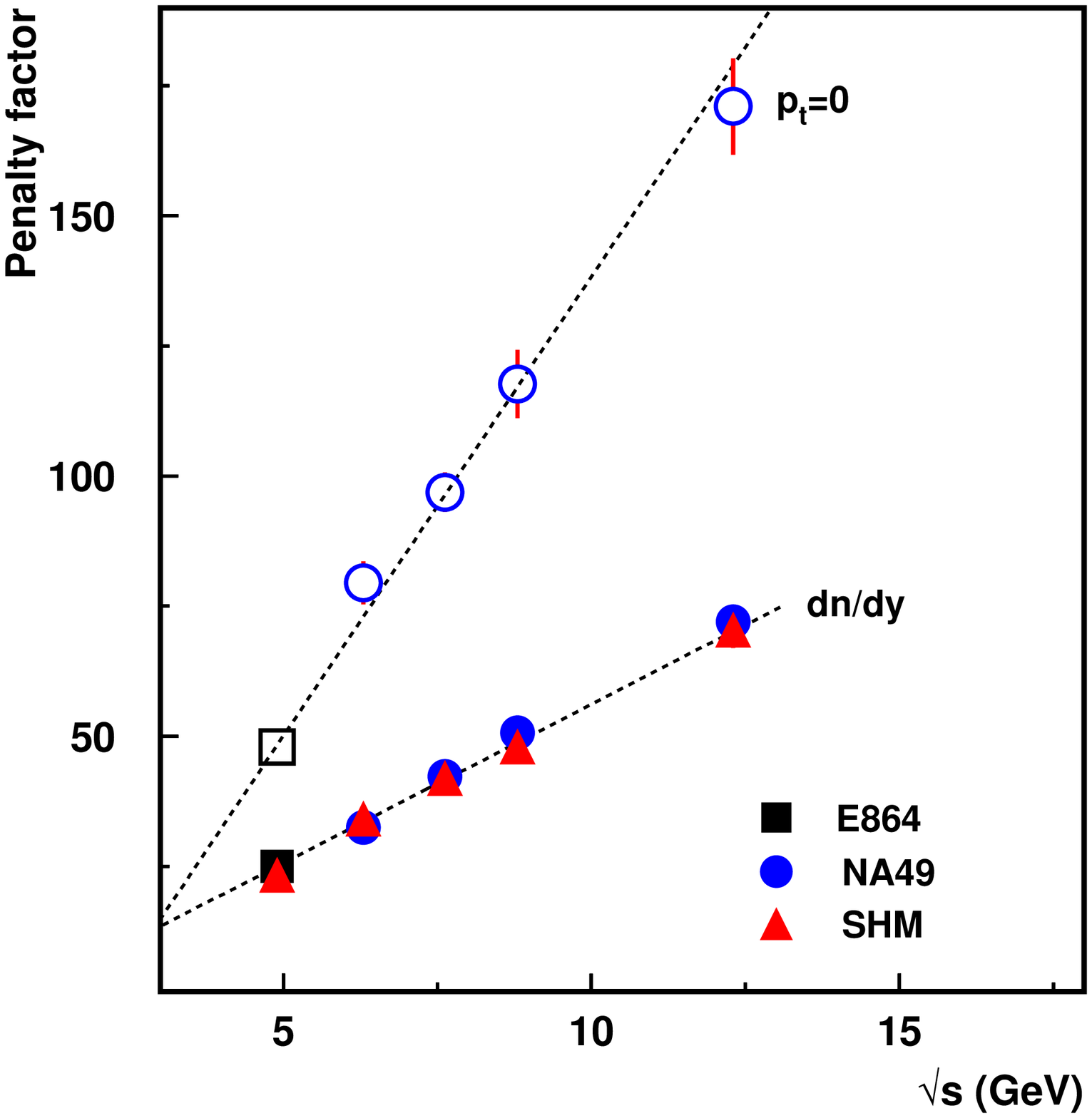}
\end{center}
\end{minipage}
\begin{minipage}[b]{45mm}
\begin{center}
\includegraphics[height=44mm]{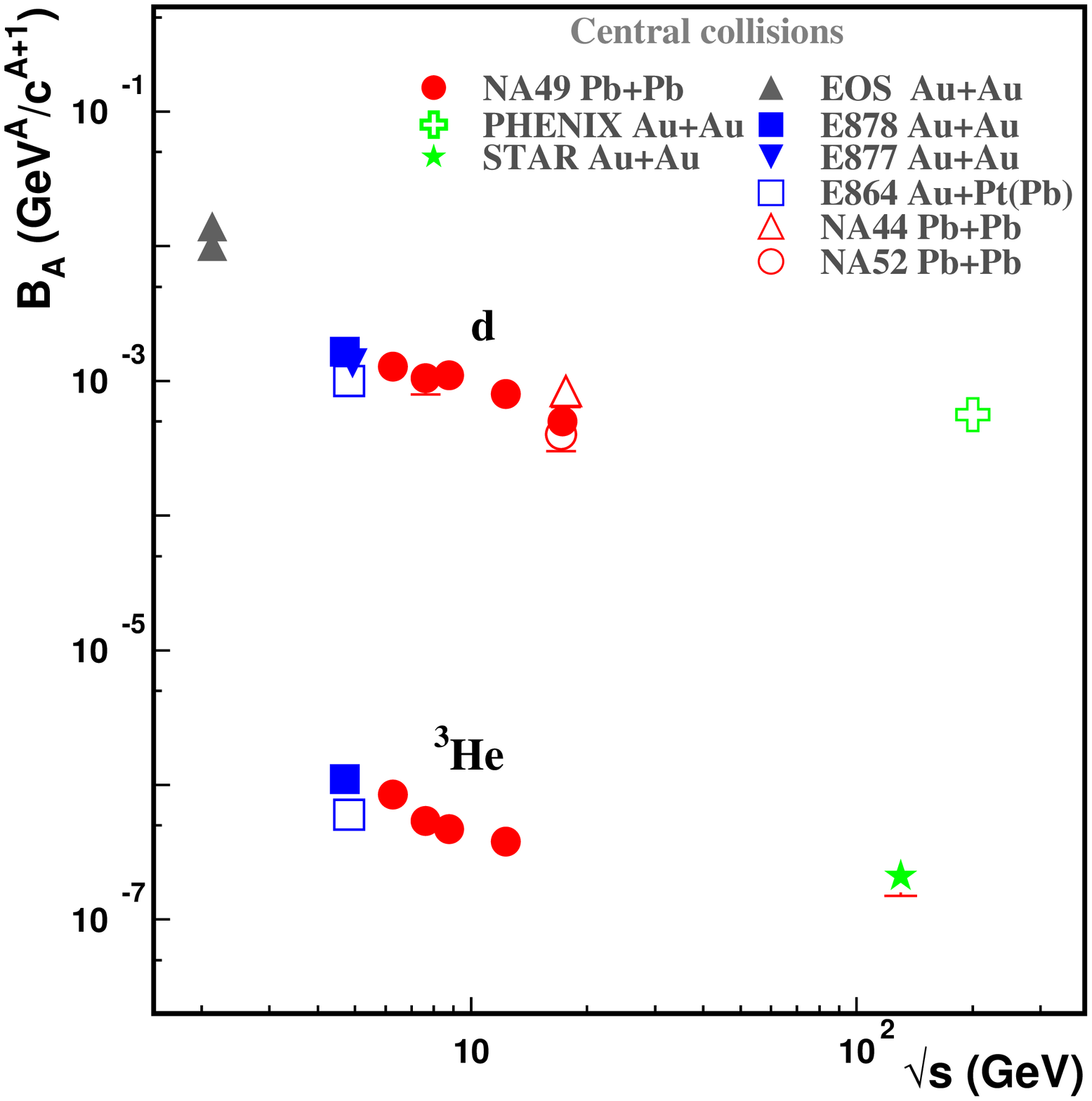}
\end{center}
\end{minipage}
\end{center}
\caption{
(left) Total multiplicity of $^3He$ in central Pb+Pb
collisions as a function of \sqrts\ (solid points).  The triangles
represent predictions of a statistical hadron gas model \cite{becc2007}.
(center) Penalty factor derived from the midrapidity $^3He$, $d$, $p$ 
production yields in central Pb+Pb collisions (dots) for $p_t$=0 and 
$p_t$-integrated yields compared to predictions of the statistical 
hadron gas model \cite{becc2007} (triangles).
(right) Energy dependence of coalescence parameters $B_2$ and $B_3$ 
at $p_t=0$ for $d$ and $^3He$, respectively.
}
\label{fig5}
\end{figure}

The total yields of $^3He$ have been estimated by fitting and integrating a parabolic 
parameterisation up to beam rapidity. 
These values shown in Fig.~\ref{fig4}~(left) agree very well 
with a prediction of a statistical hadron gas model \cite{becc2007}.
It is an overlooked aspect of the thermal model to compute the yields of composite 
particles reproduced with the chemical freeze-out parameters used to describe baryon 
and meson ratios \cite{pbm2002}.   
In these investigations, an observed exponential decrease of composite particles 
implies a penalty factor for each edditional nucleon.
In the relevant Boltzmann approximation this penalty factor, 
$R_p\approx exp \frac{m \pm \mu_b}{T}$ 
, where $m$ is a nucleon mass, can be easily derived.
Using the chemical freeze-out parameters $\mu_b$ and $T$ appropriate for SPS energies 
obtained from the statistical model fit \cite{becat2004} to the NA49 data, the penalty 
factors were calculated as well as the yields of $^3He$. 
Fig.~\ref{fig4}~(center) illustrates the results for the penalty factor $R_p$
as determined from $dn_A/dy = const / R_{p}^{A-1}$ for AGS and SPS energies 
using the yields of $p$, $d$, and $^3He$. 
The values increase linearly with 
$\sqrt{s_{NN}}$ and are largest for $p_t=0$ yields. 
For $p_t$-integrated midrapidity yields $dn/dy$ the penalty factors decrease 
because of transverse flow. One finds their values to be in surprising agreement 
with the statistical hadron gas model predictions \cite{becc2007}. 
As was mentioned above this agreement also holds for total $^3He$ yields. 
Both these observations might imply that $^3He$ is already formed at the chemical 
freeze-out, which might be in contradiction to the assumption of cluster formation 
via coalescence at possibly lower thermal freeze-out temperatures.

Using a general prescription of coalescence model which relates the invariant yield 
of light nuclei with mass $A$ to the $A$th power of the proton yield \cite{llope1995},
assuming the neuteron and proton distributions are identical:

\begin{equation}
E_A\frac{d^3N_A}{dp_A^3}= B_A\left(E_p\frac{d^3N_p}{dp_p^3}\right)^A,  p_A=Ap_p,
\end{equation}
the coalescence parameters $B_2$ and $B_3$ were extracted for deuterons and $^3He$, 
respectively. Their values at $p_t$ measured by NA49 are shown in Fig.~\ref{fig5}~(right) 
and compared to results from lower and higher energies. One observes a gradual decrease
with increasing beam energy which would suggest an increasing effective coalescence volume 
in the coalescence model scenario.

\begin{figure}[htb]
\begin{center}
\begin{minipage}[b]{45mm}
\begin{center}
\includegraphics[height=48mm]{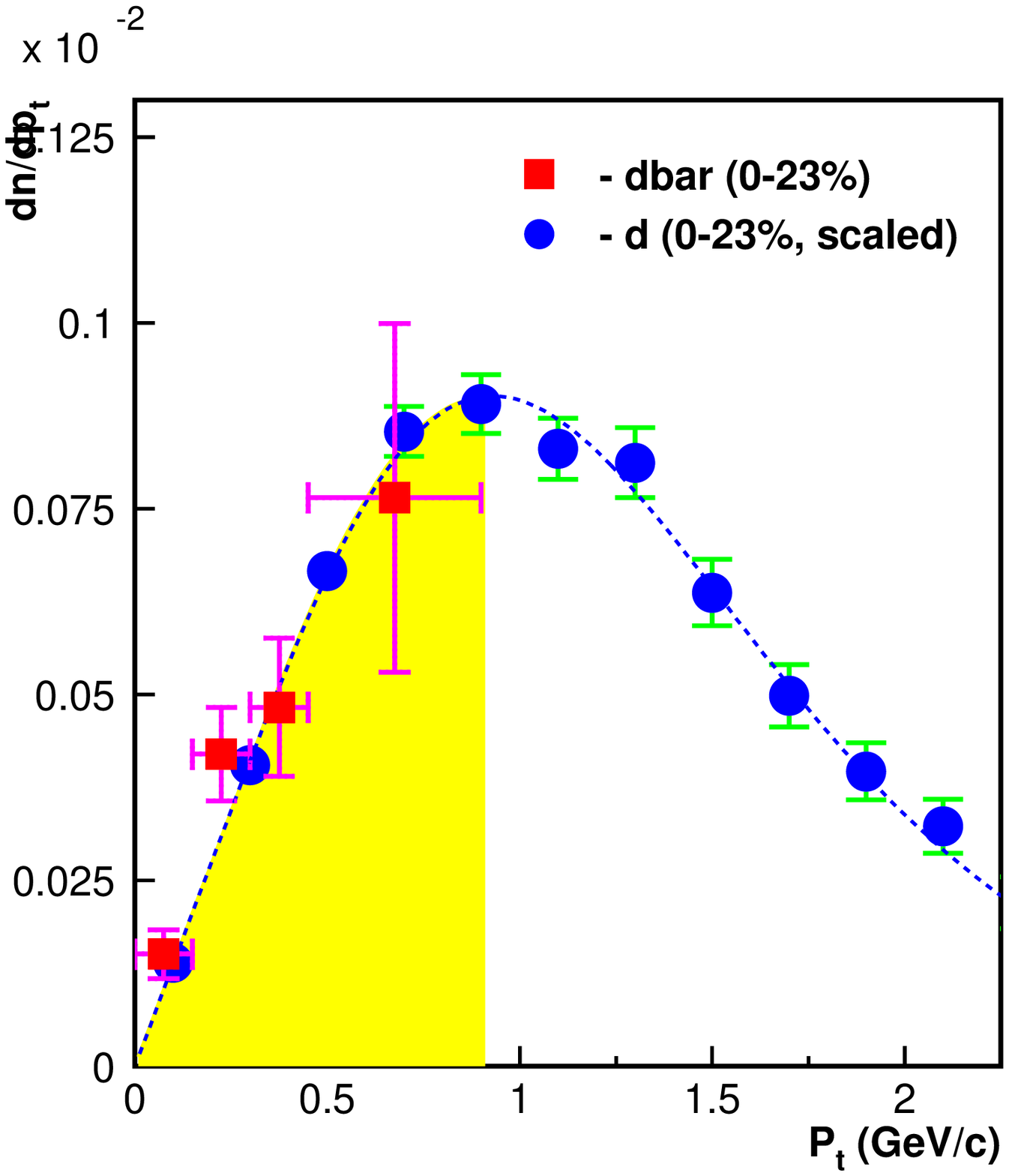}
\end{center}
\end{minipage}
\begin{minipage}[b]{45mm}
\begin{center}
\includegraphics[height=45mm]{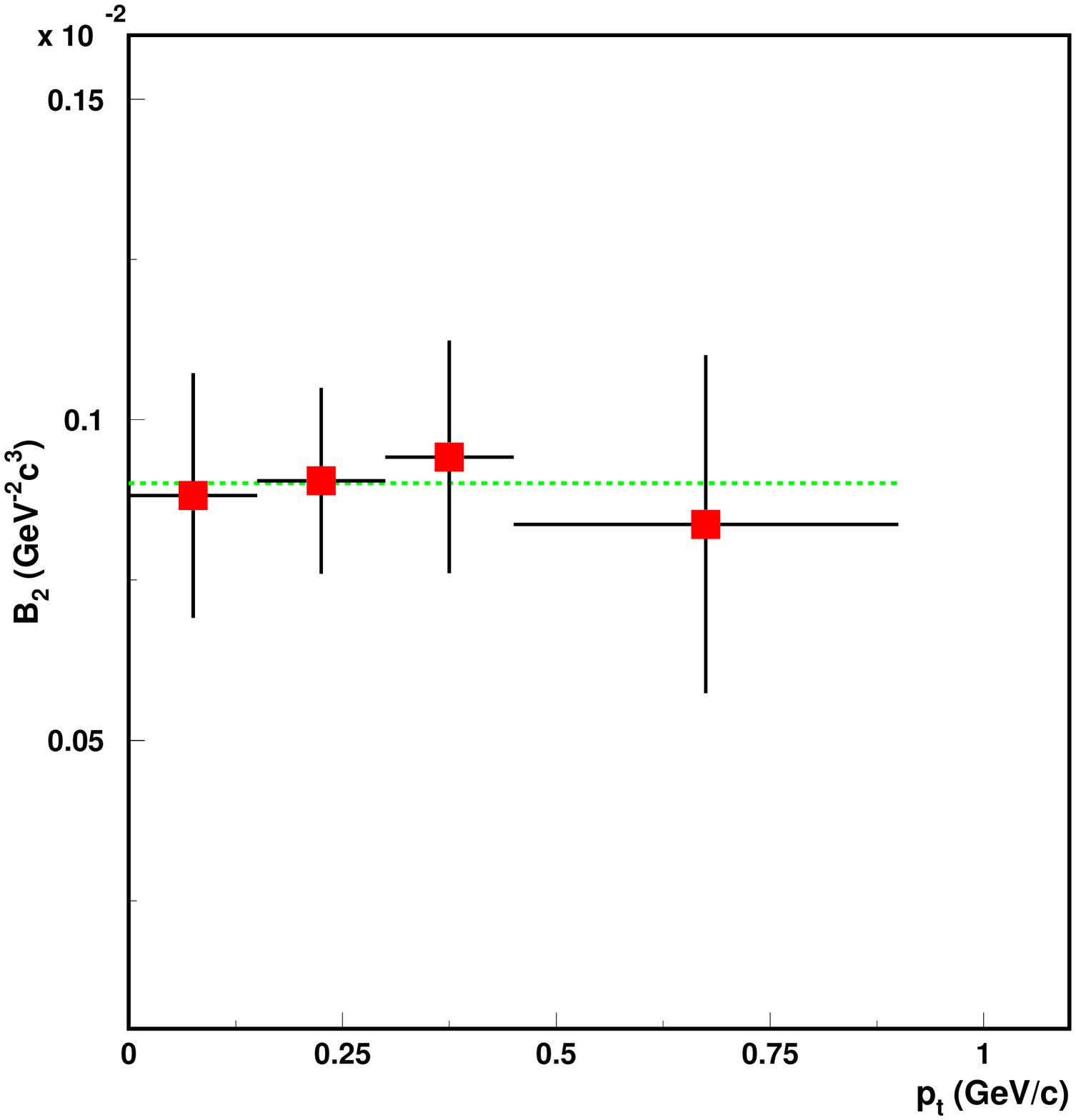}
\end{center}
\end{minipage}
\begin{minipage}[b]{45mm}
\begin{center}
\includegraphics[height=45mm]{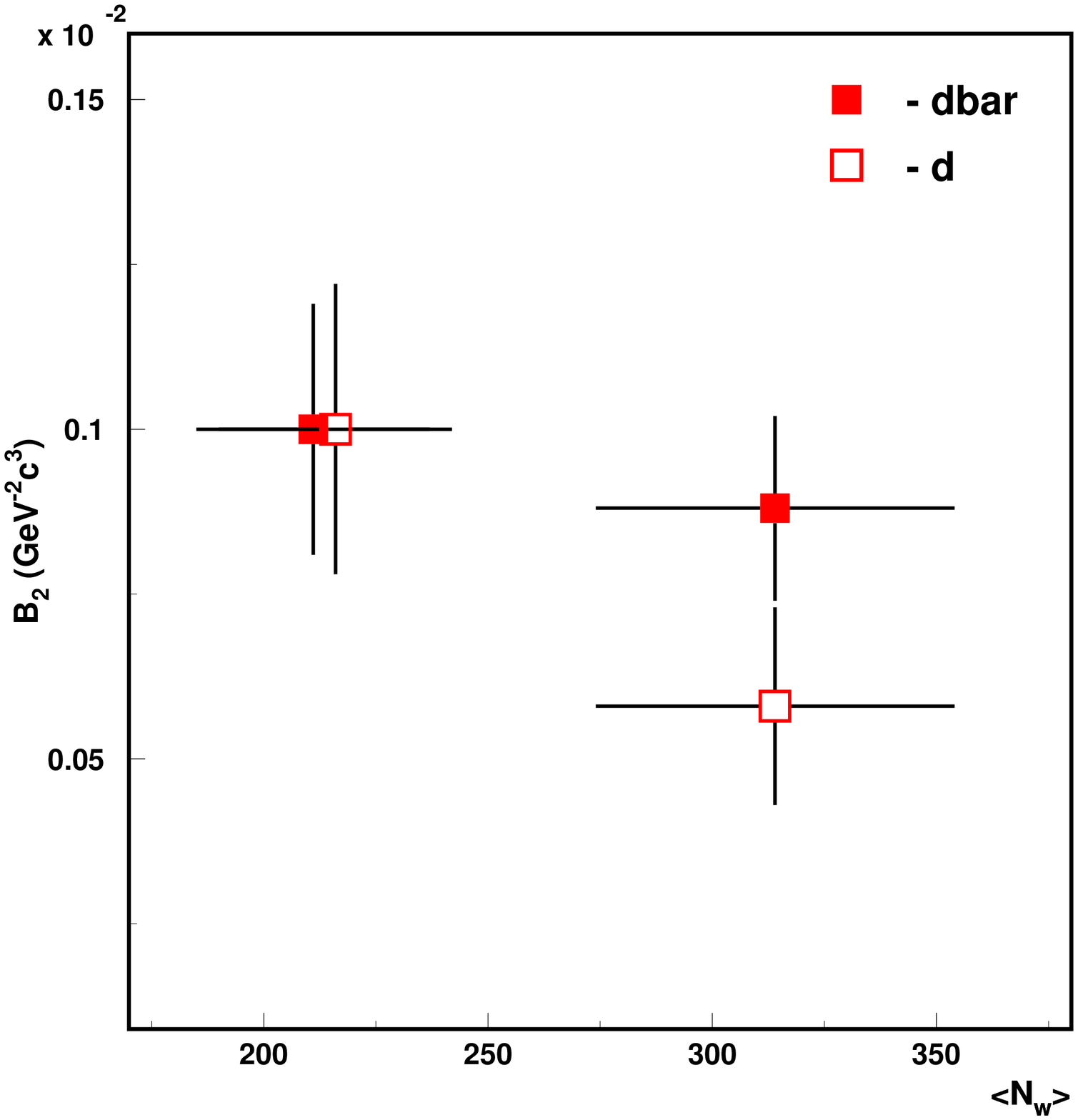}
\end{center}
\end{minipage}
\end{center}
\caption{
(left) Transverse momentum distributions for antideuterons (0<$p_t$<0.9 
GeV/$c^2$) and that scaled for deuterons (0<$p_t$<2.0) GeV/$c^2$) in  
158$A$ GeV Pb+Pb collisions at (0-23)\% centrality.
Coalescence parameter $B_2$ for antideuterons as a function of $p_t$ 
(center) and the number of wounded nucleons $\langle N_{w} \rangle$ (right).
}
\label{fig6}
\end{figure}

Analysis of antideuteron production was provided using 2.6 million events recorded 
in 158A GeV Pb+Pb collisions. 
For data set, a trigger requiring events from the 23\% most central events was used 
by measuring the total energy of spectator 
nucleons and fragments in the veto calorimeter downstream of the target.
Further off-line selection of the data for two classes of events corresponding 
to (0-10)\% and (10-23)\% centralities was used in analysis.
The measurements were performed in rapidity range $2.0<y<2.5$ and transverse momentum 
 $0<p_t<0.9$ GeV/$c$. 
The yield of antideuterons as a function of $p_t$ is presented in Fig.~\ref{fig6}~(left) 
together with that for deuterons scaled for comparison. 
A quite similar shapes of the transverse momentum distributions for $\bar{d}$ 
and $d$ in the common range of $p_t$ is observed from the figure. 
The ratio $\bar{d}/d \approx (2.0 \pm{0.2}) \cdot{10}^{-3}$ was calculated. 

The coalescence parameter for antideuterons $B_2(\bar{d})$ 
for the 23\% most central Pb+Pb events as a function of $p_t$ is plotted 
in  Fig.~\ref{fig6}~(center).
As seen from the data the $B_2(\bar{d})$ is almost independent on $p_t$ 
and its average value is calculated $B_2(\bar{d})=(9.1 \pm 0.9)\cdot10^{-4}$ 
GeV$^2$/$c^3$.
The coalescence parameters for $\bar{d}$ and $d$ as a function of the number of 
wounded nucleons for two selected centrality classes 
are shown in Fig.~\ref{fig6}~(right). Being equal within errors $B_2$ for both 
$\bar{d}$ and $d$ increases with centrality of 158$A$ GeV Pb+Pb collisions. 
This trend is usually attributed to the developed radial flow and strong transverse 
expansion of the system created in relativistic heavy ion rections.

\section{Hyperon production}\label{hyperons}

The study of strange particle production in relativistic nucleus-nucleus collisions 
plays an important role because the strangeness enhancement is one of the primary 
signatures predicted for possible creation of the quark-gluon plasma.   
 
New NA49 results on the \lam\ and \lab\ production in minimum bias Pb+Pb reactions 
at 40$A$ and 158\agev at midrapidity ($|y| < 0.4$ for \lam\ (\lab) and $|y| < 0.5$ 
for \xim) together with preliminary data on \xim\ \cite{mitr2006} are presented.  
The midrapidity transverse mass distributions for these particles in each of the 
selected centality bins were obtained and the inverse slope parameters were 
calculated from the Boltzmann exponential fits to the data:
 
\begin{equation} 
\frac{d^2n}{m_t dm_t } = C \cdot exp^{-(m_t-m)/T}.
\label{eq5}
\end{equation}

\begin{figure}[htb]
\begin{center}
\begin{minipage}[b]{60mm}
\begin{center}
\includegraphics[height=55mm]{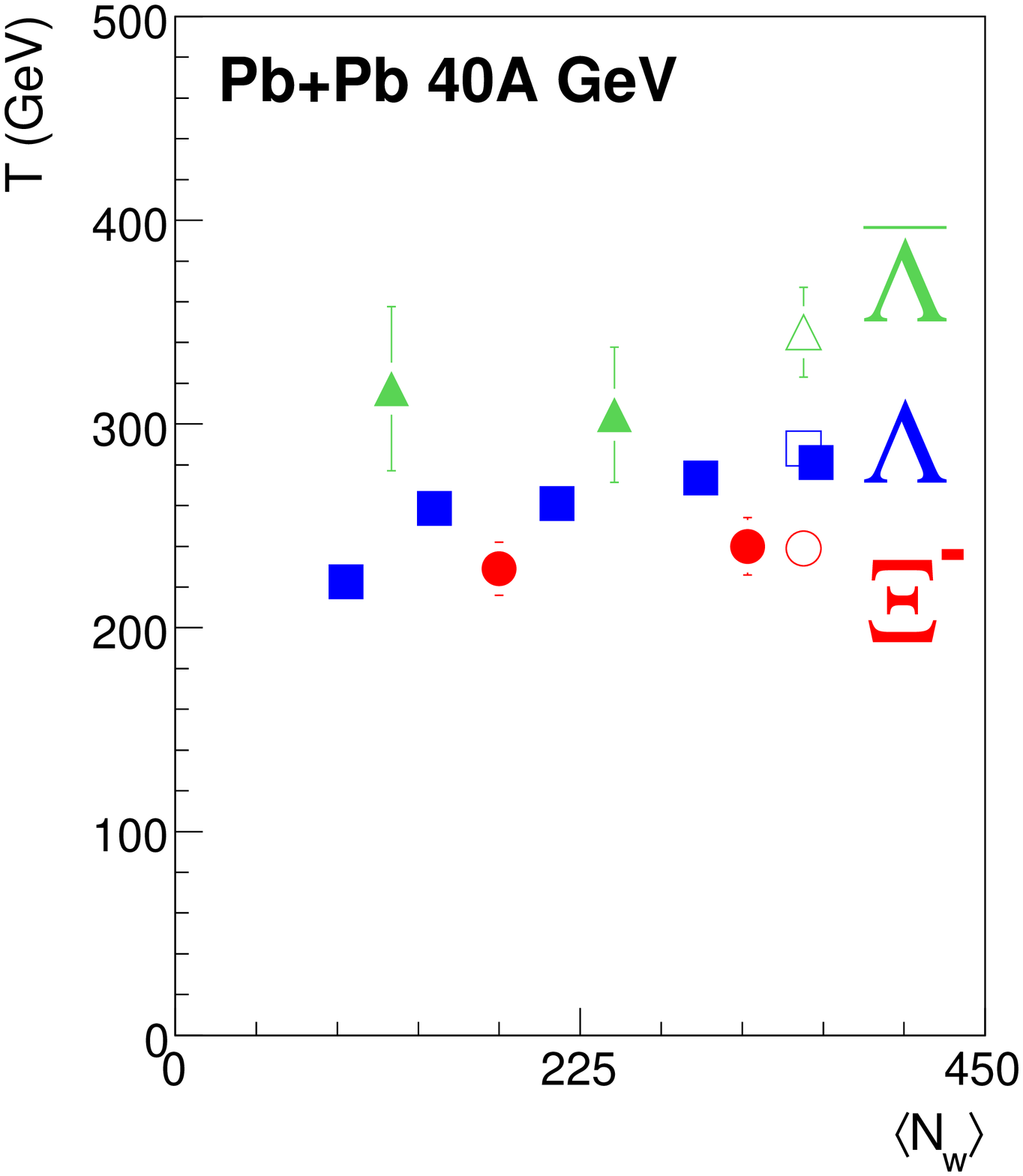}
\end{center}
\end{minipage}
\begin{minipage}[b]{60mm}
\begin{center}
\includegraphics[height=55mm]{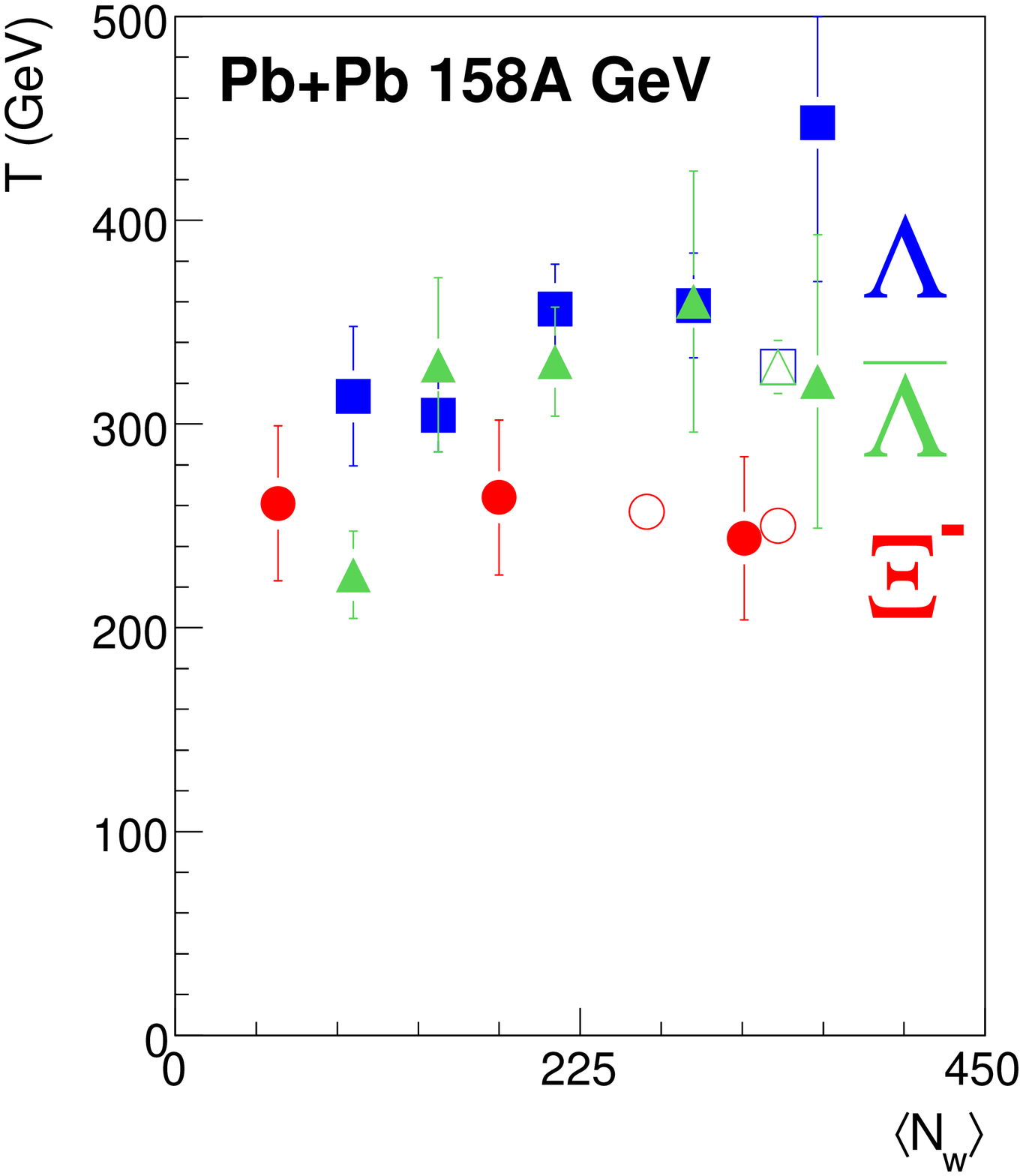}
\end{center}
\end{minipage}
\end{center}
\caption{
Inverse slope parameters for $\Lambda$, $\bar{\Lambda}$ 
and $\Xi^-$ at midrapidity for minimum bias Pb+Pb collisions (filled symbols) 
at 40$A$ GeV (left) and 158$A$ GeV (right) as a function of the number of 
wounded nucleons $\langle N_{w} \rangle$. 
Open symbols represents the results for online selected central reactions. 
}
\label{fig7}
\end{figure}

The centrality dependence of the inverse slope parameters for 40$A$ and 158$A$ 
GeV  Pb+Pb collisions is depicted in Fig.~\ref{fig7}. It is determined in terms 
of the number of wounded nucleons $\langle N_{w} \rangle$ calculated within the 
Glauber model \cite{glaub1970}. 
It is seen from the figures that the inverse slopes of $\Lambda$ and $\bar{\Lambda}$
slightly increase when going from peripheral to central Pb+Pb collisions, whereas 
no centrality (system size) dependence is observed for $\Xi^-$ cascade. The latter
might be attributed to the early freeze-out of multistrange baryons $\Xi$ and $\Omega$ 
discussed in \cite{nuxu1999}. 

\begin{figure}[htb]
\begin{minipage}[b]{50mm}
\begin{center}
\includegraphics[height=50mm]{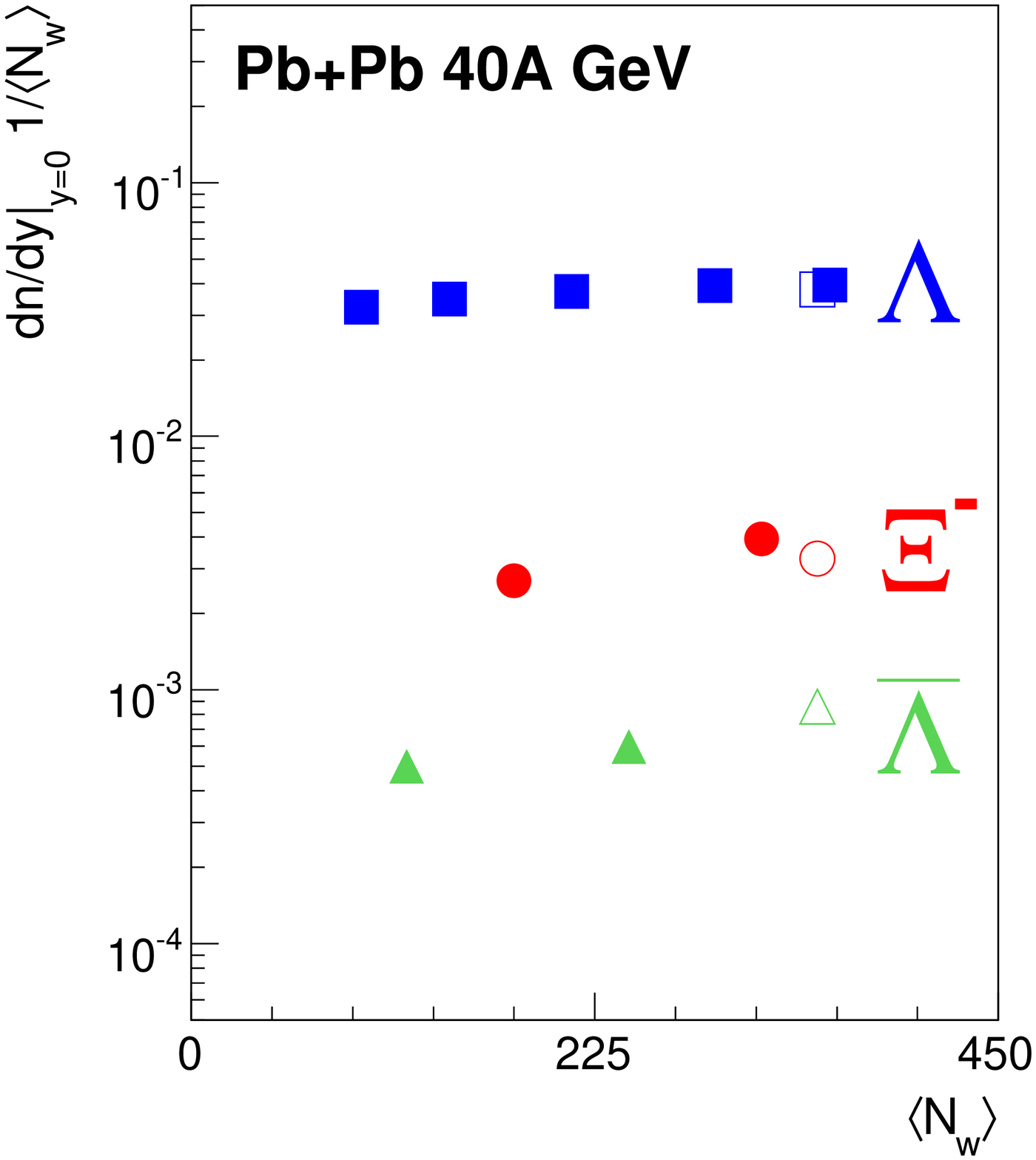}
\end{center}
\vspace{1mm}
\end{minipage}
\begin{minipage}[b]{47mm}
\begin{center}
\includegraphics[height=50mm]{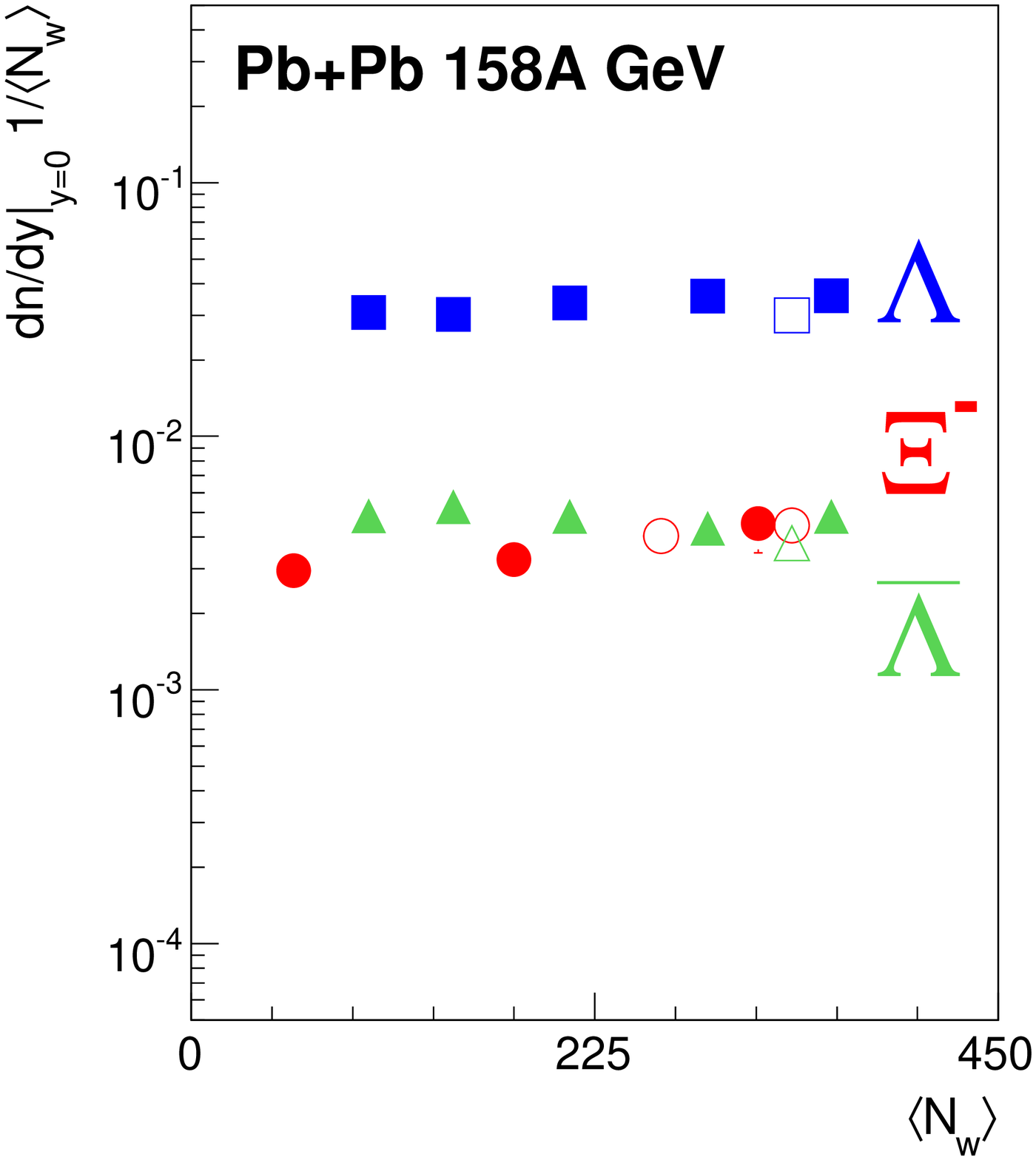}
\end{center}
\vspace{1mm}
\end{minipage}
\begin{minipage}[b]{47mm}
\begin{center}
\includegraphics[height=56mm]{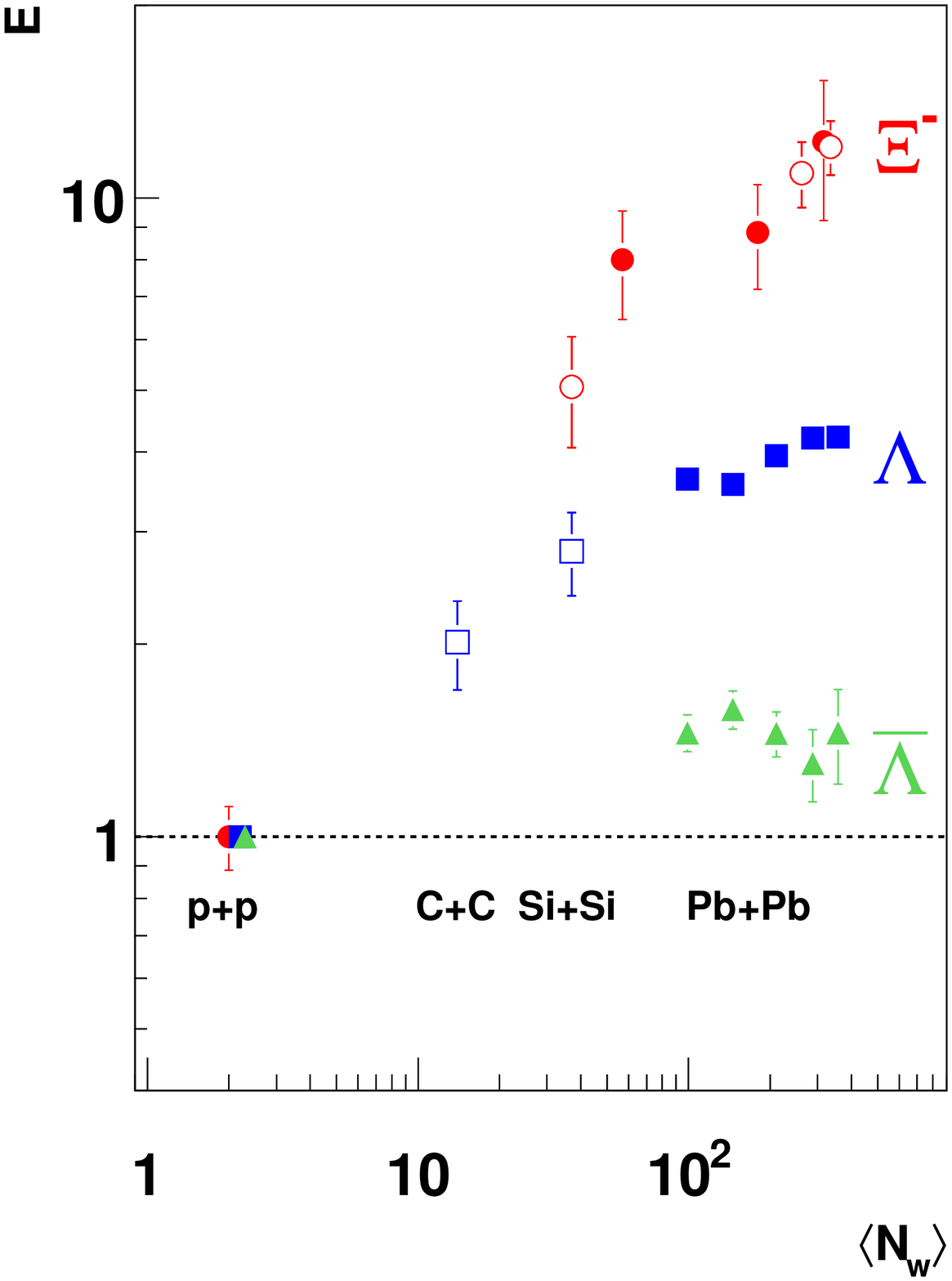}
\end{center}
\end{minipage}
\caption{
Rapidity densities per wounded nucleon of \lam, \lab, and \xim\ at midrapidity 
for minimum bias Pb+Pb collisions (filled symbols) at  40\agev\ (left) and 
158\agev\ (center) as a function of the number of wounded nucleons \nwound.  
Open symbols represent the results for online selected central reactions.    
(right) The midrapidity yields per wounded nucleon relative to p+p yields for 
central C+C, Si+Si and minimum bias Pb+Pb reactions at 158\agev as a function 
of \nwound.
}
\label{fig8}
\end{figure}

The centality dependence of hyperon yield (the rapidity density) per wounded 
nucleon 
for $\Lambda$, $\bar{\Lambda}$ and $\Xi^-$ for 40$A$ and 158$A$ GeV is shown in 
Fig.~\ref{fig8}~(left and center). The yields were calculated from the measured 
transverse mass spectra and using a fitted function extrapolated to the full 
$m_t$ range. 
The \lam\ data are corrected for feed down from weak decays. 
While there is no system size dependence of the rapidity densities per wounded 
nucleon for \lam\ and \lab, a weak rise can be observed in the case of the \xim.

Having measured by NA49 the particle yields for p+p, central C+C and Si+Si \cite{altPRL2005}, 
and minimum bias Pb+Pb collisions at 158$A$ GeV a hyperon enhancement factor $E$ can be determined as 
a ratio of the particle yields in A+A to p+p collisions:

\begin{equation}
E = \left(\frac{dn/dy}{\langle N_w \rangle}\right)_{AA}
/\left(\frac{dn/dy}{2}\right)_{pp}  
\end{equation}
A behaviour of the enhancements with centrality of the collisions is displayed 
in Fig.~\ref{fig8}~(right), where the data for selected 
centrality classes plotted as a function of the number of wounded nucleons.
A clear hierarchy of the enhancement is visible. They increase with  
strangeness content of the particles ($E(\xim) > E(\lam) > E(\lab)$). 
For \xim\ and \lam\ a moderate \nwound\ dependence is seen while for 
the $\bar{\Lambda}$ it is compatible with being flat with centrality.  
About similar trends have been observed by the NA57 collaboration, however 
relative to a p+Be baseline \cite{antin2006}.  
The difference between $\Lambda$ and $\bar{\Lambda}$ is unclear. This could be due 
to a difference in the initial production mechanism or a centrality-dependent 
$\bar{\Lambda}$ absorption in a baryon-rich environment.

\section{Conclusions}\label{concl}

New results were presented on the energy dependence of baryon stopping, 
production of light (anti)nuclei $^3He$, $t$, $\bar{d}$ and $d$ and 
centrality dependence of (anti)strange baryons \lam, \lab, and \xim in 
Pb+Pb collisions throughout the CERN SPS energy range. 

A clear evolution with energy from a peak to a dip structure is 
seen in the net baryon distributions within 20$A$-158$A$ GeV energy range.    
The relative net baryon rapidity shift  
$\langle \delta y \rangle$/$y_{proj}$$\approx{0.6}$
at the AGS and SPS and slowly decreases towards RHIC energies.
The estimated inelasticity coefficient indicate that average energy 
loss of projectiles amounts to about (70-80)\%.
 
The quantitative results on the energy dependence of total multiplicity of $^3He$ 
and penalty factor derived from $^3He$, $d$ and $p$ data are in remarkable agreement 
with statistical hadron gas model predictions. 
The measured yields of the light nuclei $\bar{d}$ and $d$ together with those for 
$\bar{p}$ and $p$ at midrapidity and nonzero $p_t$ in the 23\% most central 
Pb+Pb collisions at 158$A$ GeV has allowed to extract and compare a coalescence 
parameters $B_2(\bar{d})$ and $B_2(d)$. 
Their values turned out to be almost similar with no visible dependence on the 
transverse momentum as expected from the coalescence models. 

The hyperon yields per wounded nucleon at 158$A$ GeV Pb+Pb collisions are enhanced 
with respect to p+p interactions. A visible centrality dependence of 
the strangeness enhancement is observed for $\Lambda$ and $\Xi^-$ and not 
for $\bar{\Lambda}$ hyperons. 
The inverse slope chracteristics of the measured hyperon $m_t$ distributions at
various centralities could support the conception of early freeze-out of multistrange 
hyperons. 

\section*{Acknowledgments}

Sincere thanks to the organisers for a stimulating workshop
and for the opportunity to show results from NA49.

This work was supported by the US Department of Energy Grant 
DE-FG03-97ER41020/A000, the Bundesministerium fur Bildung und Forschung, 
Germany, 
the Virtual Institute VI-146 of Helmholtz Gemeinschaft, Germany,
the Polish State Committee for Scientific Research (1 P03B 006 30, 1 
P03B 097 29
, 1 PO3B 121 29, 1 P03B 127 30),
the Hungarian Scientific Research Foundation (T032648, T032293, 
T043514),
the Hungarian National Science Foundation, OTKA, (F034707),
the Polish-German Foundation, the Korea Science \& Engineering 
Foundation (R01-2
005-000-10334-0),
the Bulgarian National Science Fund (Ph-09/05) and the Croatian Ministry 
of Science, Education and Sport (Project 098-0982887-2878).


\begin{thebibliography}{99}

\bibitem{gazd2004} 
M.~Gazdzicki, \emph{Report from NA49}, \emph{J.Phys.} {\bf G30}, S701 (2004).

\bibitem{zak2006} 
P.~Seyboth, \emph{Onset of deconfinement in Pb+Pb collisions at the CERN SPS}, 
\emph{Acta Phys. Pol} {\bf B37}, 3429 (2006).

\bibitem{afan1999} 
S.~Afanasiev et al., \emph{The NA49 large acceptance hadron detector}, 
\emph{Nucl. Instrum. Meth.} {\bf A430}, 210 (1999).

\bibitem{afan2002} 
S.~Afanasiev et al., \emph{Energy dependence of pion and kaon production 
in central Pb + Pb collisions} \emph{Phys.~Rev.~C} {\bf 66} (2002) 054902.

\bibitem{ppbarPRC2006} 
C.~Alt et al., \emph{Energy and centrality dependence of $\bar{p}$ and $p$ 
production and the $\bar{\Lambda}$/$\bar{p}$ ratio in 
Pb+Pb collisions between 20$A$ GeV and 158$A$ GeV}, \emph{Phys. 
Rev.} {\bf C73}, 044910 (2006).

\bibitem{anti2004} 
T.~Anticic et al., \emph{$\Lambda$ and $\bar{\Lambda}$ production in central 
Pb+Pb collisions at 40, 80 and 158$A$ GeV}, \emph{Phys.~Rev.~Lett.} {\bf 93}, 
022302 (2004).

\bibitem{alt2005} 
C.~Alt et al., \emph{$\Omega$ and $\bar{\Omega}$ production in central Pb+Pb 
collisions at 40$A$ GeV and 158$A$ GeV}, \emph{Phys.~Rev.~Lett.} {\bf 94}, 
192301 (2005).

\bibitem{appel1999} 
H.~Appelsh\"{a}user, \emph{Baryon Stopping and Charged Particle Distributions 
in Central Pb+Pb Collisions at 158 GeV/$c$ per Nucleon}, \emph{Phys.Rev.Lett.} 
{\bf 82}, 2471 (1999).

\bibitem{ahle1999} 
L.~Ahle et al., \emph{Proton and deuteron production in Au+Au reactions at 11.6
GeV/$c$}, \emph{Phys. Rev.} {\bf C60}, 064901 (1999).

\bibitem{beard2004} 
I.~G.~Bearden et al., \emph{Nuclear stopping in Au+Au collisions 
at $\sqrt{s_{NN}}$=200 GeV}, \emph{Phys. Rev. Lett.} {\bf 93}, 102301 (2004).

\bibitem{urqmd} 
H.~Weber et al., \emph{Nucleus-Nucleus collisions at high baryon densities}, 
\emph{Phys.~Lett.} {\bf B545}, 285 (2002). 

\bibitem{becc2007}
F.~Becattini et al., \emph{Energy and system size dependence of chemical 
freeze-out in relativistic nuclear collisions}, \emph{Phys. Rev.} 
{\bf C73}, 044905 (2006) and private communication.

\bibitem{poll1998} 
A.~Polleri et al., \emph{Effect of collective expansion on light cluster 
spectra in relativistic heavy ion collisions}, \emph{Phys. Lett.} {\bf B419}, 
19 (1998).

\bibitem{heinz1999} 
R.~Scheibl and U.~Heinz, \emph{Coalescence and flow in ultrarelativistic 
heavy ion collisions}, \emph{Phys. Rev.} {\bf C59}, 1585 (1999).

\bibitem{dpPRC2004}
T.~Anticic et al., \emph{Energy and centrality dependence of deuteron and 
proton production in Pb+Pb collisions at relativistic energies}, \emph{Phys. 
Rev.} {\bf C69}, 024902 (2004).

\bibitem{mattiello1997} 
R.~Mattiello et al., \emph{Nuclear clusters as a probe for expansion flow 
in heavy ion reactions at (10-15)$A$ GeV}, \emph{Phys. Rev.} {\bf C55}, 1443 (1997).

\bibitem{hansen1999} 
A.~Hansen, \emph{Light nuclei production in ultrarelativistic heavy ion 
collisions}, \emph{PhD thesis, The Niels Bohr Institute.} (August 1999).

\bibitem{rqmd} 
M.~J.~Bennett et al., \emph{Light nuclei production in relativistic 
Au+nucleus collisions}, \emph{Phys. Rev.} {\bf C58}, 1155 (1998).

\bibitem{pbm2002} 
P.~Braun-Munzinger and J.~Stachel, \emph{Particle ratios, equilibration and 
the QCD phase boundary}, \emph{J.Phys.} {\bf G28}, 1971-1976 (2002).

\bibitem{becat2004}
F.~Becattini et al., \emph{Chemical equilibrium study in nucleus-nucleus 
collisions at relativistic energies}, \emph{Phys. Rev.} {\bf C69}, 024905 (2004).

\bibitem{llope1995} 
V.~J.~Llope et al., \emph{The fragment coalescence model}, \emph{Phys. 
Rev.} {\bf C52}, 2004 (1995).

\bibitem{mitr2006} 
M.~Mitrovski et al., \emph{Strangeness production at SPS energies}, 
\emph{J. Phys.} {\bf G32}, S43 (2006). 

\bibitem{glaub1970} 
R.~J.~Glauber and G.~Matthiae et al., \emph{High-energy scattering of 
protons by nuclei}, \emph{Nucl. Phys.} {\bf B21}, 135 (1970).

\bibitem{nuxu1999} 
H.~van~Hecke. et al., \emph{Evidence of early multi-strange hadron freeze-out 
in high energy nuclear collisions}, \emph{Nucl.Phys.} {\bf A661} (1999) 493c-496c.

\bibitem{altPRL2005} 
C.~Alt et al., \emph{System-size dependence od strangeness production in 
nucleus-nucleus collisions at $\sqrt{s_{NN}}/2$=17.3 GeV}, 
\emph{Phys. Rev. Lett.} {\bf 94}, 052301 (2005).

\bibitem{antin2006} 
F.~Antinori, \emph{Enhancement of hyperon production at central rapidity in
158 $A$ CeV/$c$ Pb-Pb collisions}, \emph{J.Phys.} {\bf G32}, 427 (2006).



\end{thebibliography}
\end{document}